\documentstyle{mn}

\include{epsf}

\title[Ionizing Flux at High Redshifts]{Evolution of the Ionizing Background at High Redshifts}
\author[A. J. Cooke et al.]
{Andrew J. Cooke$^1$, Brian Espey$^2$ and Bob Carswell$^3$\\
$^1$Institute for Astronomy, Edinburgh, Scotland.\\
$^2$The John Hopkins University, Baltimore, USA.\\
$^3$Institute of Astronomy, Cambridge, England.}

\em\date{Submitted to MNRAS 18th December 1995.  Accepted 22nd August 1996.}

\begin{document}
\maketitle

\newcommand{\lya}{Lyman--$\alpha$}
\newcommand{\laa}{Lyman--$\alpha$ absorption}
\newcommand{\laf}{Lyman--$\alpha$ forest}
\newcommand{\etal}{et~al.}
\newcommand{\col}{\hbox{$\log( N )$}}
\newcommand{\cm}{\hbox{\rm cm}}
\newcommand{\kms}{\hbox{\rm km\s s$^{-1}$}}
\newcommand{\s}{\thinspace}
\newcommand{\ten}[1]{$\times$10$^{#1}$}
\newcommand{\flux}{ergs Hz$^{-1}$ s$^{-1}$}
\newcommand{\oflux}{\hbox{ergs cm$^{-2}$ Hz$^{-1}$ s$^{-1}$}}
\newcommand{\olflux}{\hbox{ergs cm$^{-2}$ \AA$^{-1}$ s$^{-1}$}}
\newcommand{\lflux}{log$_{10}$ ergs cm$^{-2}$ Hz$^{-1}$ sr$^{-1}$}
\newcommand{\jflux}{J$_{23}$}
\newcommand{\elim}[2]{$+#1\atop-#2$}

\begin{abstract}

The decrease in number density of \lya\ clouds near the background
quasar is an observational result which is often called the
`proximity' or `inverse' effect.  It is thought that, for nearby
clouds, the quasar's flux dominates the background radiation field,
increasing the ionization state of the clouds and reducing the
(observed) H~I column density.

In this paper we analyse a sample of 11 quasars from the literature
for which accurate column density estimates of the \lya\ lines exist.
We confirm, to a significance level of more than 3 standard
deviations, that the proximity effect exists.  If it is related to the
background flux then the intensity and evolution of the background
have been constrained.

Using a maximum likelihood method, we determine the strength of the
extragalactic ionizing background for $2.0 < z < 4.5$, taking account
of possible systematic errors in our determination and estimating the
effect of biases inherent in the data.  If the background is constant
we find that it has an intensity of  100\elim{50}{30}~\jflux, where
\jflux\ is defined as 10$^{-23}$~ergs~cm$^{-2}$~Hz$^{-1}$~sr$^{-1}$.
There is no significant evidence for a change in this value with
redshift.

\end{abstract}


%
%

\section{Introduction}
\label{sec:intro}

%
%

\subsection{The Lyman--alpha forest} 
\label{sec:laf}

Spectroscopic observations towards quasars show a large number of
intervening absorption systems.  This `forest' of lines is numerically
dominated by systems showing only the Lyman--alpha transition ---
these absorbers are called Lyman alpha clouds.

Earlier work suggests that the clouds are large, highly ionized
structures, either pressure confined (eg. Ostriker \& Ikeuchi 1983) or within
cold dark matter structures (eg. Miralda--Escud\'e \& Rees 1993; Cen, Miralda--Escud\'{e} \& Ostriker 1994;
Petitjean \& M\"{u}cket 1995; Zhang, Anninos \& Norman 1995).  However, alternative models do
exist: cold, pressure confined clouds (eg. Barcons \& Fabian 1987, but see
Rauch et~al. 1993); various shock mechanisms (Vishniac \& Bust 1987,
Hogan 1987).

Low and medium resolution spectroscopic studies of the forest
generally measure the redshift and equivalent width of each cloud.  At
higher resolutions it is possible to measure the redshift ($z$), H~I
column density ($N$, atoms per cm$^2$) and Doppler parameter ($b$,
\kms).  These are obtained by fitting a Voigt profile to the data
(Rybicki \& Lightman 1979).

Using a list of $N$, $z$ and $b$ measurements, and their associated
error estimates, the number density of the population can be studied.
Most work has assumed that the density is a separable function of $z$,
$N$ and $b$ (Rauch et~al. 1993).

There is a local decrease in cloud numbers near the background quasar
which is normally attributed to the additional ionizing flux in that
region.  While this may not be the only reason for the depletion (the
environment near quasars may be different in other respects; the cloud
redshifts may reflect systematic motions) it is expected for the
standard physical models wherever the ionising flux from the quasar
exceeds, or is comparable to, the background.

Since the generally accepted cloud models are both optically thin to
ionizing radiation and highly ionized, it is possible to correct
column densities from the observed values to those that would be seen
if the quasar were more remote.  The simplest correction assumes that
the shape of the two incident spectra --- quasar and background ---
are similar.  In this case the column density of the neutral fraction
is inversely proportional to the incident ionizing flux.

If the flux from the quasar is known, and the depletion of clouds is
measured from observations, the background flux can be determined.  By
observing absorption towards quasars at different redshifts the
evolution of the flux can be measured.  Bechtold (1993) summarises
earlier measurements of the ionising flux, both locally and at higher
redshifts.

Recently Loeb \& Eisenstein (1995) have suggested that enhanced clustering near
quasars causes this approach to overestimate the background flux.  If
this is the case then an analysis which can also study the evolution
of the effect gives important information.  In particular, a decrease
in the inferred flux might be expected after the redshift where the
quasar population appears to decrease.  However, if the postulated
clustering enhancement is related to the turn--on of quasars at high
redshift, it may conspire to mask any change in the ionizing
background.

Section \ref{sec:model} describes the model of the population density
in more detail, including the corrections to flux and redshift that
are necessary for a reliable result.  The data used are described in
section \ref{sec:data}.  In section \ref{sec:errors} the quality of
the fit is assessed and the procedure used to calculate errors in the
derived parameters is explained.  Results are given in section
\ref{sec:results} and their implications discussed in section
\ref{sec:discuss}.  Section \ref{sec:conc} concludes the paper.

%
%

\section{The Model}
\label{sec:model}

%
%

\subsection{Population Density}
\label{sec:popden}

The Doppler parameter distribution is not included in the model since
it is not needed to determine the ionizing background from the proximity
effect.  The model here assumes that $N$ and $z$ are uncorrelated.
While this is unlikely (Carswell 1995), it should be a good
approximation over the restricted range of column densities considered
here.

The model of the population without Doppler parameters or the
correction for the proximity effect is 
\begin{equation} dn(N^\prime,z) = \, A^\prime
(1+z)^{\gamma^\prime}\,(N^\prime)^{-\beta}
\frac{c(1+z)}{H_0(1+2q_0z)^\frac{1}{2}} \ dN^\prime\,dz
\end{equation}
where $H_0$ is the Hubble parameter, $q_0$ is the cosmological
deceleration parameter and $c$ is the speed of light.  Correcting for
the ionizing flux and changing from `original' ($N^\prime$) to
`observed' ($N$) column densities, gives 
\begin{equation} dn(N,z) = \, A (1+z)^{\gamma^\prime}
\left(\frac{N}{\Delta_F}\right)^{-\beta}
\frac{c(1+z)}{H_0(1+2q_0z)^\frac{1}{2}} \ \frac{dN}{\Delta_F}\,dz
\end{equation}
where
\begin{equation}N = N^\prime \Delta_F \ , \end{equation}
\begin{equation}\Delta_F = \frac{ f_\nu^B }{ f_\nu^B + f_\nu^Q } \ , 
\end{equation}
and $f_\nu^B$ is the background flux, $f_\nu^Q$ is the flux from the
quasar ($4\pi J_\nu(z)$).

The background flux $J_\nu^B$ may vary with redshift.  Here it is
parameterised as a constant, a power law, or two power laws with a
break which is fixed at $z_B=3.25$ (the mid--point of the available
data range).  An attempt was made to fit models with $z_B$ as a free
parameter, but the models were too poorly constrained by the data to
be useful.

\begin{eqnarray}
J_\nu(z)=10^{J_{3.25}} & \hbox{model {\bf B}} \\
J_\nu(z)=10^{J_{3.25}}\left(\frac{1+z}{1+3.25}\right)^{\alpha_1} & \mbox{{\bf C}} \\
J_\nu(z)=10^{J_{z_B}}\times\left\{\begin{array}{ll}\left(\frac{1+z}{1+z_B}\right)^{\alpha_1}&\mbox{$z<z_B$}\\
\left(\frac{1+z}{1+z_B}\right)^{\alpha_2}&\mbox{$z>z_B$}\end{array}\right. & \mbox{{\bf D \& E}}
\end{eqnarray}

A large amount of information (figure \ref{fig:nz}) is used to
constrain the model parameters.  The high--resolution line lists give
the column density and redshift, with associated errors, for each
line.  To calculate the background ionising flux the quasar luminosity
and redshift must be known (table \ref{tab:objects}).  Finally, each
set of lines must have observational completeness limits (table
\ref{tab:compl}).

%
%

\subsection{Malmquist Bias and Line Blending}
\label{sec:malm}

Malmquist bias is a common problem when fitting models to a population
which increases rapidly at some point (often near an observational
limit).  Errors during the observations scatter lines away from the
more populated regions of parameter space and into less populated
areas.  Line blending occurs when, especially at high redshifts,
nearby, overlapping lines cannot be individually resolved.  This is a
consequence of the natural line width of the clouds and cannot be
corrected with improved spectrographic resolution.  The end result is
that weaker lines are not detected in the resultant `blend'.  Both
these effects mean that the observed population is not identical to
the `underlying' or `real' distribution.

\subsubsection{The Idea of Data Quality}

To calculate a correction for Malmquist bias we need to understand the
significance of the error estimate since any correction involves
understanding what would happen if the `same' error occurs for
different column density clouds.  The same physical cloud cannot be
observed with completely different parameters, but the same
combination of all the complex factors which influence the errors
might affect a line with different parameters in a predictable way.
If this idea of the `quality' of an observation could be quantified it
would be possible to correct for Malmquist bias: rather than the
`underlying' population, one reflecting the quality of the observation
(ie. including the bias due to observational errors) could be fitted
to the data.

For example, if the `quality' of an observation was such that,
whatever the actual column density measured, the error in column
density was the same, then it would be trivial to convolve the
`underlying' model with a Gaussian of the correct width to arrive at
an `observed' model.  Fitting the latter to the data would give
parameters unaffected by Malmquist bias.  Another example is the case
of galaxy magnitudes.  The error in a measured magnitude is a fairly
simple function of source brightness, exposure time, etc., and so it
is possible to correct a flux--limited galaxy survey for Malmquist
bias.

\subsubsection{Using Errors as a Measure of Quality}

It may be possible to describe the quality of a spectrum by the signal
to noise level in each bin.  From this one could, for a given line,
calculate the expected error in the equivalent width.  The error in
the equivalent width might translate, depending on whether the
absorption line was in the linear or logarithmic portion of the `curve
of growth', to a normal error in either $N$ or $\log(N)$.  But in this
idealised case it has been assumed that the spectrum has not been
re--binned, leaving the errors uncorrelated; that the effect of
overlapping, blended lines is unimportant; that there is a sudden
transition from a linear to logarithmic curve of growth; that the
resulting error is well described by a normal distribution.  None of
this is likely to be correct and, in any case, the resulting analysis,
with different `observed' populations for every line, would be too
unwieldy to implement, given current computing facilities.

\begin{figure}
\epsfxsize=8.5cm
\epsfbox{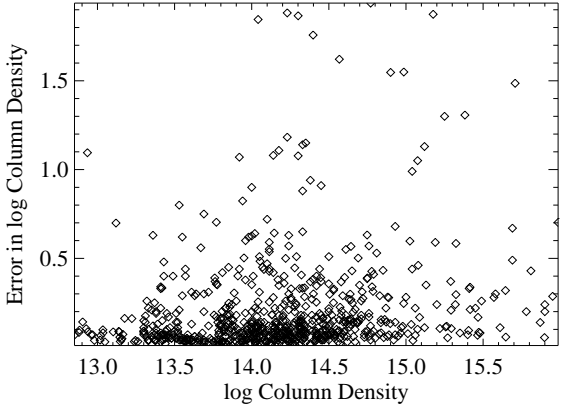}
\epsfverbosetrue
\caption{The distribution of errors in column density.} 
\label{fig:ndn}
\end{figure}

A more pragmatic approach might be possible.  A plot of the
distribution of errors with column density (figure~\ref{fig:ndn})
suggests that the errors in $\log(N)$ are of a similar magnitude for a
wide range of lines (although there is a significant correlation
between the two parameters).  Could the error in $\log(N)$ be a
sufficiently good indicator of the `quality' of an observation?

If the number density of the underlying population is $n^\prime(N)\
\hbox{d}\log(N)$ then the observed population density for a line with
error in $\log(N)$ of $\sigma_N$ is: 
\begin{equation} n(N)\ \hbox{d}\log(N)
\propto\int_{-\infty}^{\infty}n^\prime\left(N10^x\right)\,\exp\left(\frac{-x^2}{2\sigma_N^2}\right)\,\hbox{d}x\
.  \end{equation}
For a power law distribution this can be calculated analytically
and gives an increased probability of seeing lines with larger errors,
as expected.  For an underlying population density $N^{-\beta}\
\hbox{d}N$ the increase is $\exp\left((1-\beta)^2(\sigma_N\log
10)^2/2\right)$.

This gives a lower statistical weight to lines with larger errors when
fitting.  For this case --- a power law and log--normal errors --- the
weighting is not a function of $N$ directly, which might imply that
any correction would be uniform, with little bias expected for
estimated parameters.

In practice this correction does not work.  This is probably because
the exponential dependence of the correction on $\sigma_N$ makes it
extremely sensitive to the assumptions made in the derivation above.
These assumptions are not correct.  For example, it seems that the
correlation between \col\ and the associated error is important.

\subsubsection{An Estimation of the Malmquist Bias}

It is possible to do a simple numerical simulation to gauge the
magnitude of the effect of Malmquist bias.  A population of ten
million column densities were selected at random from a power law
distribution ($\beta=1.5, \log(N_{\hbox{min}})=10.9,
\log(N_{\hbox{max}})=22.5$) as an `unbiased' sample.  Each line was
given an `observed' column density by adding a random error
distributed with a normal or log--normal (for lines where $13.8 <
\log(N) < 17.8$) distribution, with a mean of zero and a standard
deviation in $\log(N)$ of $0.5$.  This procedure is a simple
approximation to the type of errors coming from the curve of growth
analysis discussed above, assuming that errors are approximately
constant in $\log(N)$ (figure~\ref{fig:ndn}).  The size of the error
is larger than typical, so any inferred change in $\beta$ should be an
upper limit.

Since a power--law distribution diverges as $N\rightarrow0$ a normal
distribution of errors in $N$ would give an infinite number of
observed lines at every column density.  This is clearly unphysical
(presumably the errors are not as extended as in a normal distribution
and the population has some low column density limit).  Because of
this the `normal' errors above were actually constrained to lie within
3 standard deviations of zero.  

\begin{figure}
\epsfxsize=8.5cm
\epsfbox{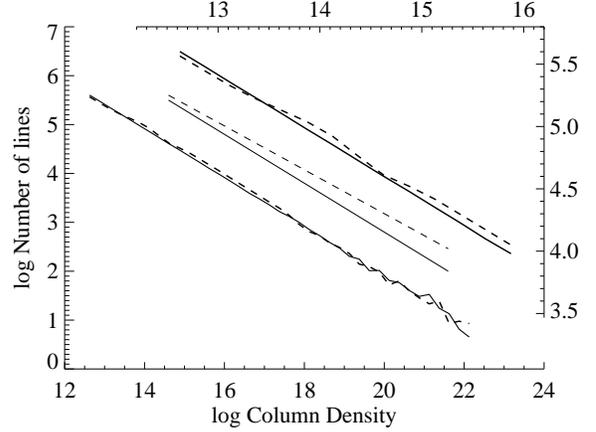}
\epsfverbosetrue
\caption{A model including Malmquist bias.  The bold, solid line is the original sample, the bold, dashed line is the distribution after processing as described in the text.  Each curve is shown twice, but the upper right plot has both axes scaled by a factor of 3 and only shows data for $12.5<\log(N)<16$.  Reference lines showing the evolution expected for $\beta=1.5$ and $1.45$ (dashed) are also shown (centre).} 
\label{fig:malm}
\end{figure}

The results (figure~\ref{fig:malm}) show that Malmquist bias has only
a small effect, at least for the model used here.  The main solid line
is the original sample, the dashed line is the observed population.
Note that the results in this paper come from fitting to a sample of
lines with $12.5<\log(N)<16$ (section~\ref{sec:data}) ---
corresponding to the data shown expanded to the upper right of the
figure.  Lines with smaller column densities are not shown since that
fraction of the population is affected by the lower density cut--off
in the synthetic data.

A comparison with the two reference lines, showing the slopes for a
population with $\beta=1.5$ or $1.45$ (dashed), indicates that the
expected change in $\beta$ is $\sim0.05$.  The population of lines
within the logarithmic region of the curve of growth appears to be
slightly enhanced, but otherwise the two curves are remarkably
similar.  The variations at large column densities are due to the
small number of strong lines in the sample.

\subsubsection{Other Approaches}

What other approaches can be used to measure or correct the effects of
Malmquist bias and line blending?  Press \& Rybicki (1993) used a completely
different analysis of the \laf.  Generating and reducing synthetic
data, with a known background and cloud population, would allow us to
assess the effect of blending.  Changing (sub--setting) the sample of
lines that is analysed will alter the relative (and, possibly,
absolute) importance of the two effects.

The procedure used by Press \& Rybicki (1993) is not affected by Malmquist bias
or line blending, but it is difficult to adapt to measure the ionizing
background.

Profile fitting to high--resolution data is a slow process, involving
significant manual intervention (we have tried to automate
profile--fitting with little success).  An accurate measurement of the
systematic error in the ionizing background would need an order of
magnitude more data than is used here to get sufficiently low error
limits.  Even if this is possible --- the analysis would need a
prohibitive amount of CPU time --- it would be sufficient work for a
separate, major paper (we would be glad to supply our software to
anyone willing to try this).

Taking a sub--set of the data is not helpful unless it is less likely
to be affected by the biases described above.  One approach might be
to reject points with large errors, or large relative errors, in
column density since these are more affected by Malmquist bias.
However, this would make the observations incomplete in a very poorly
understood way.  For example, relative errors are correlated with
column density (as noted above) and so rejecting lines with larger
relative errors would preferentially reject higher column density
lines.  There is no sense in trying to measure one bias if doing so
introduces others.

Unlike rejecting lines throughout the sample, changing the completeness
limit does not alter the coverage of the observations (or rather, it
does so in a way that is understood and corrected for within the
analysis).  Raising the completeness limits should make line blending
less important since weaker lines, which are most likely to be
blended, are excluded from the fit.  Whether it affects the Malmquist
bias depends on the distribution of errors.

For blended lines, which tend to be weak, raising the completeness
limit should increase the absolute value of $\beta$ since the more
populous region of the (hypothetical) power--law population of column
densities will no longer be artificially depleted.  The effect on
$\gamma$ is more difficult to assess since it is uncertain whether the
 completeness limits are correct at each redshift.  If the limits
increase too rapidly with redshift, for example, then raising them
further will reduce blending most at lower redshifts, lowering
$\gamma$.  But if they are increasing too slowly then the converse
will be true.

\subsubsection{Conclusions}

Until either profile--fitting is automated, or the method of
Press \& Rybicki (1993) can be modified to include the proximity effect, these
two sources of uncertainty --- Malmquist bias and line blending ---
will continue be a problem for any analysis of the \laf.  However,
from the arguments above, it is likely that the effect of Malmquist
bias is small and that, by increasing the completeness limit, we can
assess the magnitude of the effect of line blending.

%
%

\subsection{Flux Calculations}
\label{sec:fluxcal}

%
%

\subsubsection{Galactic Extinction}

Extinction within our Galaxy reduces the apparent luminosity of the
quasars and so lowers the estimate of the background.  Since the
absorption varies with frequency this also alters the observed
spectral slope.

Observed fluxes were corrected using extinction estimates derived from
the H~I measurements of Heiles \& Cleary (1979) for Q2204--573 and
Stark et~al. (1992) for all the other objects.  H I column densities were
converted to $E(B-V)$ using the relationships: \begin{eqnarray}
E(B-V)&=&\frac{N_{\hbox{\footnotesize
H~I}}}{5.27\,10^{21}}\quad\hbox{if\ } \frac{N_{\hbox{\footnotesize
H~I}}}{5.27\,10^{21}}<0.1\\ E(B-V)&=&\frac{N_{\hbox{\footnotesize
H~I}}}{4.37\,10^{21}}\quad\hbox{otherwise} \end{eqnarray} where the
first value comes from Diplas \& Savage (1994) and the second value, which
compensates for the presence of H$_2$, is the first scaled by the
ratio of the conversions given in Bohlin, Savage \& Drake (1978).  A ratio
$R=A(V)/E(B-V)$ of 3.0 (Lockman \& Savage 1995) was used and variations of
extinction with frequency, $A(\lambda)/A(V)$ were taken from
Cardelli, Clayton \& Mathis (1989).

The correction to the observed index, $\alpha_o$, of the power--law
continuum, 
\begin{equation} f_\nu\propto\nu^{-\alpha}\ , \end{equation}
was calculated using
\begin{equation}
\alpha_o=\alpha+\frac{A(V)}{2.5}\frac{\partial}{\partial\ln\nu}\frac{A(\nu)}{A(V)}
\end{equation}
which, using the notation of Cardelli, Clayton \& Mathis (1989), becomes
\begin{equation}
\alpha_o=\alpha+\frac{A(V)}{2.5\,10^6c}\,\nu\ln(10)\,\frac{\partial}{\partial
y}\left(a(x)+\frac{b(x)}{R}\right)\ .  \end{equation}

\subsubsection{Extinction in Damped Systems}
\label{sec:dampcor}

Two quasars are known to have damped absorption systems along the line
of sight (Wolfe et~al. 1995).  The extinction due to these systems is not
certain, but model {\bf E} includes the corrections listed in
table~\ref{tab:damp}.  These have been calculated using the SMC
extinction curve in Pei (1992), with a correction for the
evolution of heavy element abundances taken from Pei \& Fall (1995).  The
SMC extinction curve is most suitable for these systems since they do
not appear to have structure at 2220~\AA\ (Boiss\'{e} \& Bergeron 1988), unlike LMC
and Galactic curves.

\begin{table}
\begin{tabular}{lllll}
Object&$\log(N_{\hbox{\footnotesize H I}})$&$z_{\hbox{\footnotesize abs}}$&$A(V)$&$\Delta_\alpha$\\
Q0000--263&21.3&3.39&0.10  &0.078\\
Q2206--199&20.7&1.92&0.14  &0.10\\
$        $&20.4&2.08&0.019 &0.049\\
\end{tabular}
\caption{
The damped absorption systems and associated corrections (at 1450~\AA\ in the quasar's rest--frame) for model {\bf E}.}
\label{tab:damp}
\end{table}

%
%

\subsubsection{Absorption by Clouds near the Quasar}
\label{sec:internal}

The amount of ionizing flux from the background quasar incident on a
cloud is attenuated by all the other clouds towards the source.  If
one of the intervening clouds has a large column density this can
significantly reduce the extent of the effect of the quasar.

To correct for this the fraction of ionizing photons from the quasar
not attenuated by the intervening H~I and He~II absorption is
estimated before fitting the model.  A power--law spectrum is assumed
and the attenuation is calculated for each cloud using the
cross--sections given in Osterbrock (1989).  The ratio
$n(\hbox{He~II})/n(\hbox{H~I})$ within the clouds will depend on
several unknown factors (the true energy distribution of the ionizing
flux, cloud density, etc.), but was assumed to be 10 (Sargent et~al. 1980,
Miralda--Escud\'e 1993).

The attenuation is calculated using all the observed intervening
clouds.  This includes clouds which are not included in the main fit
because they lie outside the column density limits, or are too close
to the quasar ($\Delta z \leq 0.003$).

For most clouds ($\log(N)\sim13.5$) near enough to the quasar to
influence the calculation of the background this correction is
unimportant (less than 1\%).  However, large ($\log(N)\sim18$ or
larger) clouds attenuate the flux to near zero.  This explains why
clouds with $\Delta_f\sim1$ are apparent close to the QSO in
figure~\ref{fig:prox}.

This relatively sudden change in optical depth at $\log(N)\sim18$ is
convenient since it makes the correction insensitive to any
uncertainties in the calculation (eg. $n(\hbox{He~II})/n(\hbox{H~I})$,
the shape of the incident spectrum, absorption by heavier elements)
--- for most column densities any reasonable model is either
insignificant ($\log(N)<17$) or blocks practically all ionizing
radiation ($\log(N)>19$).

In fact, the simple correction described above is in reasonable
agreement with CLOUDY models, for even the most critical column
densities.  A model cloud with a column density of $\log(N)=13.5$ and
constant density was irradiated by an ionizing spectrum based on that
of Haardt \& Madau (1996).  Between the cloud and quasar the model included an
additional absorber (constant density, $\log(N)=18$) which modified
the quasar's spectrum.  The effect of the absorber (for a range of
heavy element abundances from pure H to primordial to 0.1 solar) on
the ionized fraction of H~I was consistent with an inferred decrease
in the quasar flux of about 80\%.  In comparison, the correction
above, using a power--law spectrum with $\alpha=1$, gave a reduction
of 60\% in the quasar flux.  These two values are in good agreement,
considering the exponential dependence on column densities and the
uncertainty in spectral shape.  At higher and lower absorber column
densities the agreement was even better, as expected.


\subsection{Redshift Corrections}
\label{sec:redcor}

Gaskell (1982) first pointed out a discrepancy between the redshifts
measured from Lyman $\alpha$ and C~IV emission, and those from lower
ionization lines (eg. Mg~II, the Balmer series).  Lower ionization
lines have a larger redshift.  If the systemic redshift of the quasar
is assumed to be that of the extended emission (Heckman et~al. 1991),
molecular emission (Barvainis et~al. 1994), or forbidden line emission
(Carswell et~al. 1991), then the low ionization lines give a better measure
of the rest--frame redshift.

Using high ionization lines gives a reduced redshift for the quasar,
implies a higher incident flux on the clouds from the quasar, and, for
the same local depletion of lines, a higher estimate of the
background.

Espey (1993) re--analysed the data in Lu, Wolfe \& Turnshek (1991), correcting
systematic errors in the quasar redshifts. The analysis also
considered corrections for optically thick and thin universes and the
differences between the background and quasar spectra, but the
dominant effect in reducing the estimate from 174 to 50~\jflux\ was the
change in the quasar redshifts.

To derive a more accurate estimate of the systemic velocity of the
quasars in our sample we made use of published redshift measurements
of low ionization lines, or measured these where spectra were
available to us. The lines used depended on the redshift and line
strengths in the data, but typically were one or more of
Mg~II$\,2798\,$\AA, O~I$\,1304\,$\AA, and C~II$\,1335\,$\AA.

When no low ionization line observations were available (Q0420--388,
Q1158--187, Q2204--573) we applied a mean correction to the high
ionization line redshifts.  These corrections are based on
measurements of the relative velocity shifts between high and low
ionization lines in a large sample of quasars (Espey \& Junkkarinen 1996).  They
find a correlation between quasar luminosity and mean velocity
difference ($\Delta_v$) with an empirical relationship given by:
\begin{equation} \Delta_v=\exp(0.66\log L_{1450}-13.72)\
\kms\end{equation} where $L_{1450}$ is the rest--frame luminosity
(\flux) of the quasar at 1450~\AA\ for $q_0=0.5$ and H$_0=100\
\kms/\hbox{Mpc}$.

%
%

\section{The Data}
\label{sec:data}

\begin{table*}
\begin{tabular}{lrrrrrrrr}
&&&\multicolumn{2}{c}{$L_\nu(1450)$}&\multicolumn{4}{c}{Typical change in $\log(\hbox{\jflux})$}\\
\hfill Object \hfill&\hfill $z$ \hfill&\hfill $\alpha$ \hfill&\hfill $q_0=0$ \hfill&$\hfill q_0=0.5$ \hfill&\hfill $z$ \hfill&\hfill $f_\nu$ \hfill&\hfill $\alpha$ \hfill&\hfill Total \hfill\\
Q0000--263 & 4.124 & 1.02 & 13.5\ten{31} &  2.8\ten{31} & $-$0.09 & $+$0.02 & $+$0.01 & $-$0.05\\
Q0014+813 & 3.398 & 0.55 & 34.0\ten{31} &  8.6\ten{31} & $-$0.19 & $+$0.33 & $+$0.21 & $+$0.36\\
Q0207--398 & 2.821 & 0.41 &  5.6\ten{31} &  1.7\ten{31} & $-$0.16 & $+$0.03 & $+$0.02 & $-$0.11\\
Q0420--388 & 3.124 & 0.38 & 10.9\ten{31} &  3.0\ten{31} & $-$0.16 & $+$0.04 & $+$0.02 & $-$0.10\\
Q1033--033 & 4.509 & 0.46 &  5.5\ten{31} &  1.0\ten{31} & $-$0.05 & $+$0.12 & $+$0.00 & $+$0.06\\
Q1100--264 & 2.152 & 0.34 & 13.8\ten{31} &  5.3\ten{31} & $-$0.42 & $+$0.19 & $+$0.11 & $-$0.13\\
Q1158--187 & 2.454 & 0.50 & 42.2\ten{31} & 14.4\ten{31} & $-$0.46 & $+$0.09 & $+$0.06 & $-$0.31\\
Q1448--232 & 2.223 & 0.61 &  9.6\ten{31} &  3.5\ten{31} & $-$0.34 & $+$0.28 & $+$0.17 & $+$0.11\\
Q2000--330 & 3.783 & 0.85 & 12.7\ten{31} &  2.9\ten{31} & $-$0.10 & $+$0.16 & $+$0.10 & $+$0.15\\
Q2204--573 & 2.731 & 0.50 & 42.8\ten{31} & 13.3\ten{31} & $-$0.35 & $+$0.06 & $+$0.03 & $-$0.25\\
Q2206--199 & 2.574 & 0.50 & 19.4\ten{31} &  6.3\ten{31} & $-$0.31 & $+$0.05 & $+$0.03 & $-$0.22\\
Mean       & 3.081 & 0.56 & 19.1\ten{31} &  5.7\ten{31} & $-$0.24 & $+$0.12 & $+$0.07 & $-$0.04\\
\end{tabular}
\caption{
The systemic redshifts, power law continuum exponents. and rest frame luminosities (\flux\ at 1450~\AA) for the quasars used.  $H_0=100$ \kms/Mpc and luminosity scales as $H_0^{-2}$.  The final four columns are an estimate of the relative effect of the various corrections in the paper (systemic redshift, correction for reddening to flux and spectral slope).}
\label{tab:objects}
\end{table*}

Objects, redshifts and fluxes are listed in table \ref{tab:objects}.
A total of 1675 lines from 11 quasar spectra were taken from the
literature.  Of these, 844 lie within the range of redshifts and
column densities listed in table \ref{tab:compl}, although the full
sample is used to correct for absorption between the quasar and
individual clouds (section~\ref{sec:internal}).  The lower column
density limits are taken from the references; upper column densities
are fixed at $\col=16$ to avoid the double power law distribution
discussed by Petitjean et~al. (1993).  Fluxes are calculated using standard
formulae, assuming a power law spectrum ($f_\nu \propto
\nu^{-\alpha}$), with corrections for reddening.  Low ionization line
redshifts were used where possible, otherwise high ionization lines
were corrected using the relation given in section~\ref{sec:redcor}.
Values of $\alpha$ uncorrected for absorption are used where possible,
corrected using the relation above.  If no $\alpha$ was available, a
value of 0.5 was assumed (Francis 1993).

References and notes on the calculations for each object follow:

\begin{description}

\item[Q0000--263] Line list from Cooke (1994).  There is some
uncertainty in the wavelength calibration for these data, but the
error ($\sim30 \kms$) is much less than the uncertainty in the quasar
redshift ($\sim900 \kms$) which is taken into account in the error
estimate (section~\ref{sec:erress}).  Redshift this paper
(section~\ref{sec:redcor}).  Flux and $\alpha$ measurements from
Sargent, Steidel \& Boksenberg (1989).

\item[Q0014+813] Line list from Rauch et~al. (1993).  Redshift this paper
(section~\ref{sec:redcor}).  Flux and $\alpha$ measurements from
Sargent, Steidel \& Boksenberg (1989).

\item[Q0207--398] Line list from Webb (1987).  Redshift (O I line)
from Wilkes (1984).  Flux and $\alpha$ measurements from
Baldwin et~al. (1995).

\item[Q0420--388] Line list from Atwood, Baldwin \& Carswell (1985).  Redshift, flux and
$\alpha$ from Osmer (1979) (flux measured from plot).  The redshifts
quoted in the literature vary significantly, so a larger error (0.01)
was used in section~\ref{sec:erress}.

\item[Q1033--033] Line list and flux from Williger et~al. (1994).  From their
data, $\alpha=0.78$, without a reddening correction.  Redshift this
paper (section~\ref{sec:redcor}).

\item[Q1100--264] Line list from Cooke (1994).  Redshift from
Espey et~al. (1989) and $\alpha$ from Tytler \& Fan (1992).  Flux measured from
Osmer \& Smith (1977).

\item[Q1158--187] Line list from Webb (1987).  Redshift from
Kunth, Sargent \& Kowal (1981).  Flux from Adam (1985).

\item[Q1448--232] Line list from Webb (1987).  Redshift from
Espey et~al. (1989).  Flux and $\alpha$ measured from Wilkes et~al. (1983),
although a wide range of values exist in the literature and so a
larger error (0.6 magnitudes in the flux) was used in
section~\ref{sec:erress}.

\item[Q2000--330] Line list from Carswell et~al. (1987).  Redshift this paper
(section~\ref{sec:redcor}).  Flux and $\alpha$ measurements from
Sargent, Steidel \& Boksenberg (1989).

\item[Q2204--573] Line list from Webb (1987).  Redshift from
Wilkes et~al. (1983).  V magnitude from Adam (1985).

\item[Q2206--199] Line list from Rauch et~al. (1993).  Redshift this paper
(section~\ref{sec:redcor}).  V magnitude from Hewitt \& Burbidge (1989).

\end{description}

\begin{table}
\begin{tabular}{cccccc}
Object&\multispan2{\hfil$N$\hfil}&\multispan2{\hfil$z$\hfil}&Number\\
name&Low&High&Low&High&of lines\\
Q0000--263& 14.00 & 16.00 & 3.1130 & 3.3104 &  62 \\ 
           &       &       & 3.4914 & 4.1210 & 101 \\ 
Q0014+813& 13.30 & 16.00 & 2.7000 & 3.3800 & 191 \\ 
Q0207--398& 13.75 & 16.00 & 2.0765 & 2.1752 &  11 \\ 
           &       &       & 2.4055 & 2.4878 &   7 \\ 
           &       &       & 2.6441 & 2.7346 &   6 \\ 
           &       &       & 2.6852 & 2.7757 &   9 \\ 
           &       &       & 2.7346 & 2.8180 &   8 \\ 
Q0420--388& 13.75 & 16.00 & 2.7200 & 3.0800 &  73 \\ 
Q1033--033& 14.00 & 16.00 & 3.7000 & 3.7710 &  16 \\ 
           &       &       & 3.7916 & 3.8944 &  21 \\ 
           &       &       & 3.9191 & 4.0301 &  24 \\ 
           &       &       & 4.0548 & 4.1412 &  25 \\ 
           &       &       & 4.1988 & 4.3139 &  30 \\ 
           &       &       & 4.3525 & 4.4490 &  23 \\ 
           &       &       & 4.4517 & 4.4780 &   2 \\ 
Q1100--264& 12.85 & 16.00 & 1.7886 & 1.8281 &   2 \\ 
           &       &       & 1.8330 & 1.8733 &   8 \\ 
           &       &       & 1.8774 & 1.9194 &  13 \\ 
           &       &       & 1.9235 & 1.9646 &   9 \\ 
           &       &       & 1.9696 & 2.0123 &  10 \\ 
           &       &       & 2.0189 & 2.0617 &   6 \\ 
           &       &       & 2.0683 & 2.1119 &  18 \\ 
Q1158--187& 13.75 & 16.00 & 2.3397 & 2.4510 &   9 \\ 
Q1448--232& 13.75 & 16.00 & 2.0847 & 2.1752 &   9 \\ 
Q2000--330& 13.75 & 16.00 & 3.3000 & 3.4255 &  23 \\ 
           &       &       & 3.4580 & 3.5390 &  15 \\ 
           &       &       & 3.5690 & 3.6440 &  18 \\ 
           &       &       & 3.6810 & 3.7450 &  11 \\ 
Q2204--573& 13.75 & 16.00 & 2.4467 & 2.5371 &  10 \\ 
           &       &       & 2.5454 & 2.6276 &  12 \\ 
           &       &       & 2.6441 & 2.7280 &   8 \\ 
Q2206--199& 13.30 & 16.00 & 2.0864 & 2.1094 &   2 \\ 
           &       &       & 2.1226 & 2.1637 &   8 \\ 
           &       &       & 2.1760 & 2.2188 &   5 \\ 
           &       &       & 2.2320 & 2.2739 &   7 \\ 
           &       &       & 2.2887 & 2.3331 &   7 \\ 
           &       &       & 2.3471 & 2.3940 &  10 \\ 
           &       &       & 2.4105 & 2.4574 &   4 \\ 
           &       &       & 2.4754 & 2.5215 &  11 \\ 
\multicolumn{2}{l}{Total: 11 quasars }&       &        &        & 844 \\
\end{tabular}
\caption{Completeness limits.}
\label{tab:compl}
\end{table}

Table~\ref{tab:objects} also gives an estimate of the relative effect
of the different corrections made here.  Each row gives the typical
change in $\log(\hbox{\jflux})$ that would be estimated using that
quasar alone, with a typical absorption cloud 2~Mpc from the quasar
($q_0=0.5, H_0=100\,\hbox{\kms}$).  The correction to obtain the
systemic redshift is not necessary for any quasar whose redshift has
been determined using low ionization lines.  In such cases the value
given is the expected change if the redshift measurement had not been
available.

Using the systematic redshift always reduces the background estimate,
while correcting for reddening always acts in the opposite sense.  The
net result, in the final column of table~\ref{tab:objects}, depends on
the relative strength of these two factors.  For most objects the
redshift correction dominates, lowering $\log(\hbox{\jflux})$ by $\sim
0.15$ (a decrease of 30\%), but for four objects the reddening is more
important (Q0014+813, the most reddened, has $B-V = 0.33$; the average
for all other objects is $0.09$).

\begin{figure*}
\hbox{
\epsfxsize=8.5cm
\epsfbox{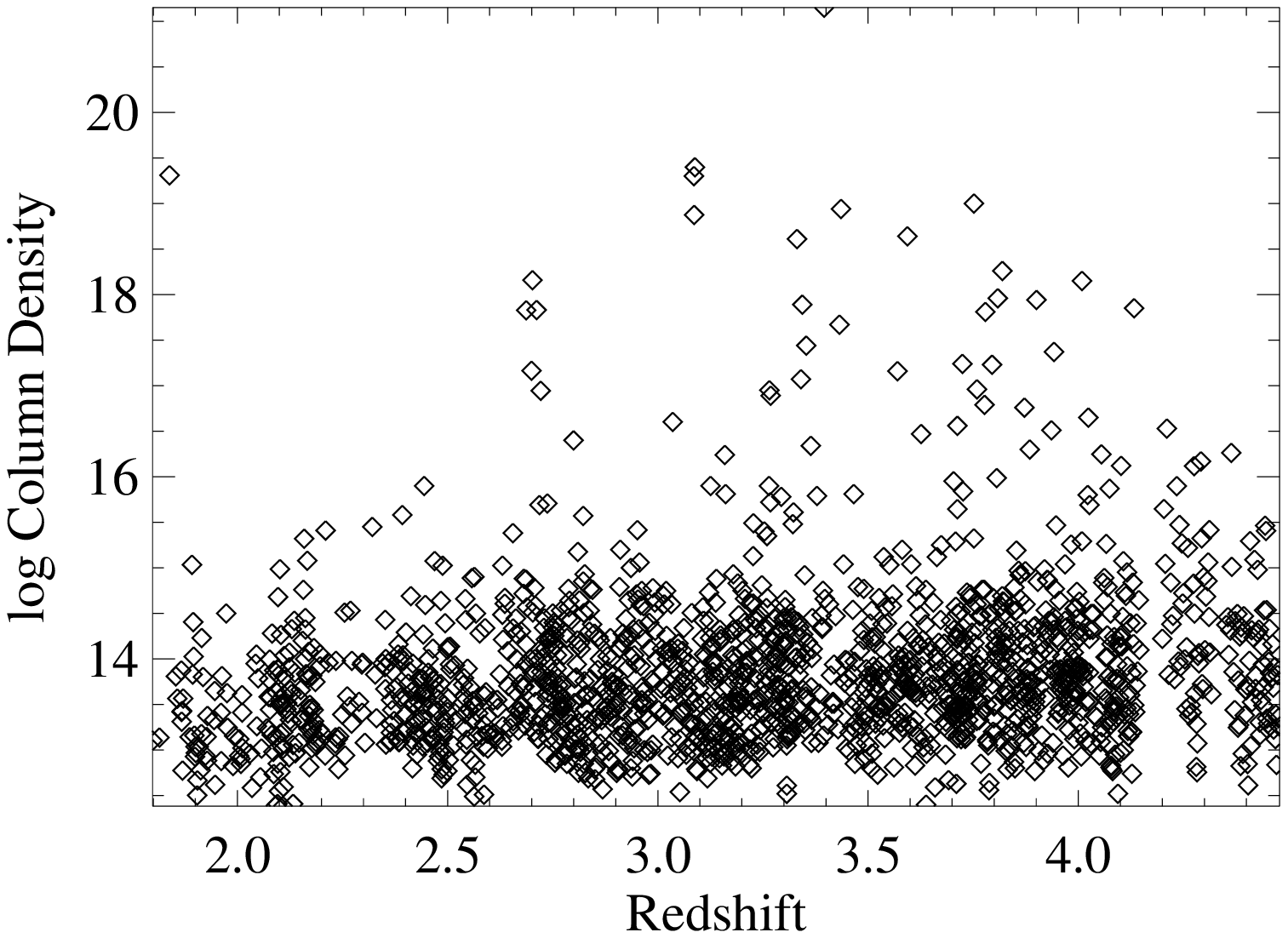}
\hfill
\epsfxsize=8.5cm
\epsfbox{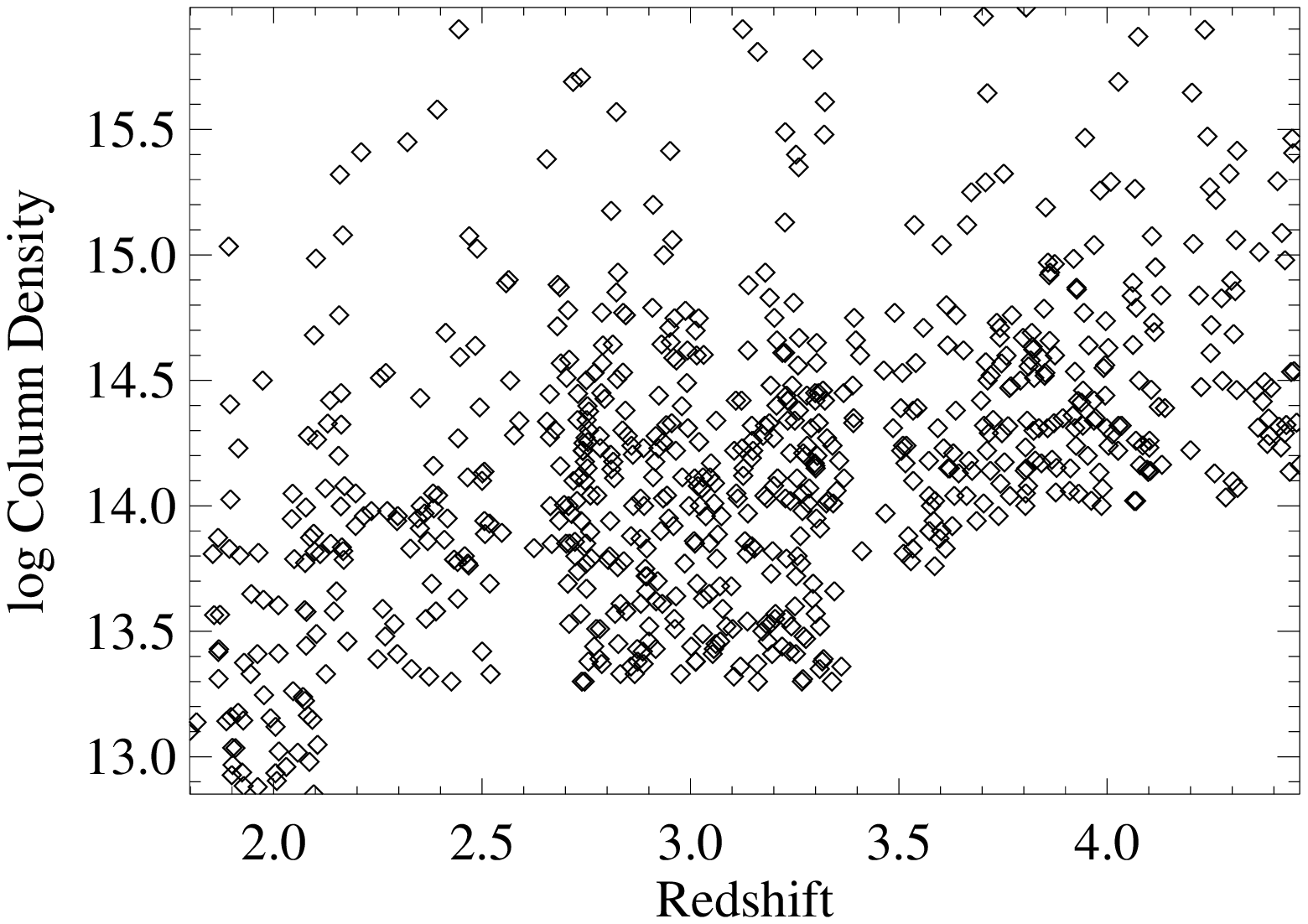}
}
\epsfverbosetrue
\caption{
The lines in the full sample (left) used to calculate the attenuation 
of the quasar flux by intervening clouds and the restricted sample 
(right) to which the model was fitted.} 
\label{fig:nz}
\end{figure*}

Figure \ref{fig:nz} shows the distribution of column density, $N$, and
redshift, $z$, for the lines in the sample.  The completeness limit
was taken from the literature and depends on the quality of the
spectra.  There is also a clear trend with redshift as the number
density increases and weak lines become less and less easy to separate
in complex blends, whatever the data quality (see
section~\ref{sec:malm} for a more detailed discussion of line
blending).

%
%

\section{Fit Quality and Error Estimates}
\label{sec:errors}

%
%

\subsection{The Quality of the Fit}
\label{sec:finalq}

\begin{table}
\begin{tabular}{c@{\hspace{3em}}cc@{\hspace{3em}}cc}
&\multicolumn{2}{c}{Without Inv. Eff.\hfill}&\multicolumn{2}{c}{With Inv. Eff.\hfill}\\
Variable&Statistic&Prob.&Statistic&Prob.\\
$N$ & 1.11 & 0.17 & 1.05 & 0.22 \\
$z$ & 1.12 & 0.16 & 1.05 & 0.22 \\
\end{tabular}
\caption{
The K--S statistics measuring the quality of the fit.}
\label{tab:ks}
\end{table}

\begin{figure*}
\hbox{
\epsfxsize=8.5cm
\epsfbox{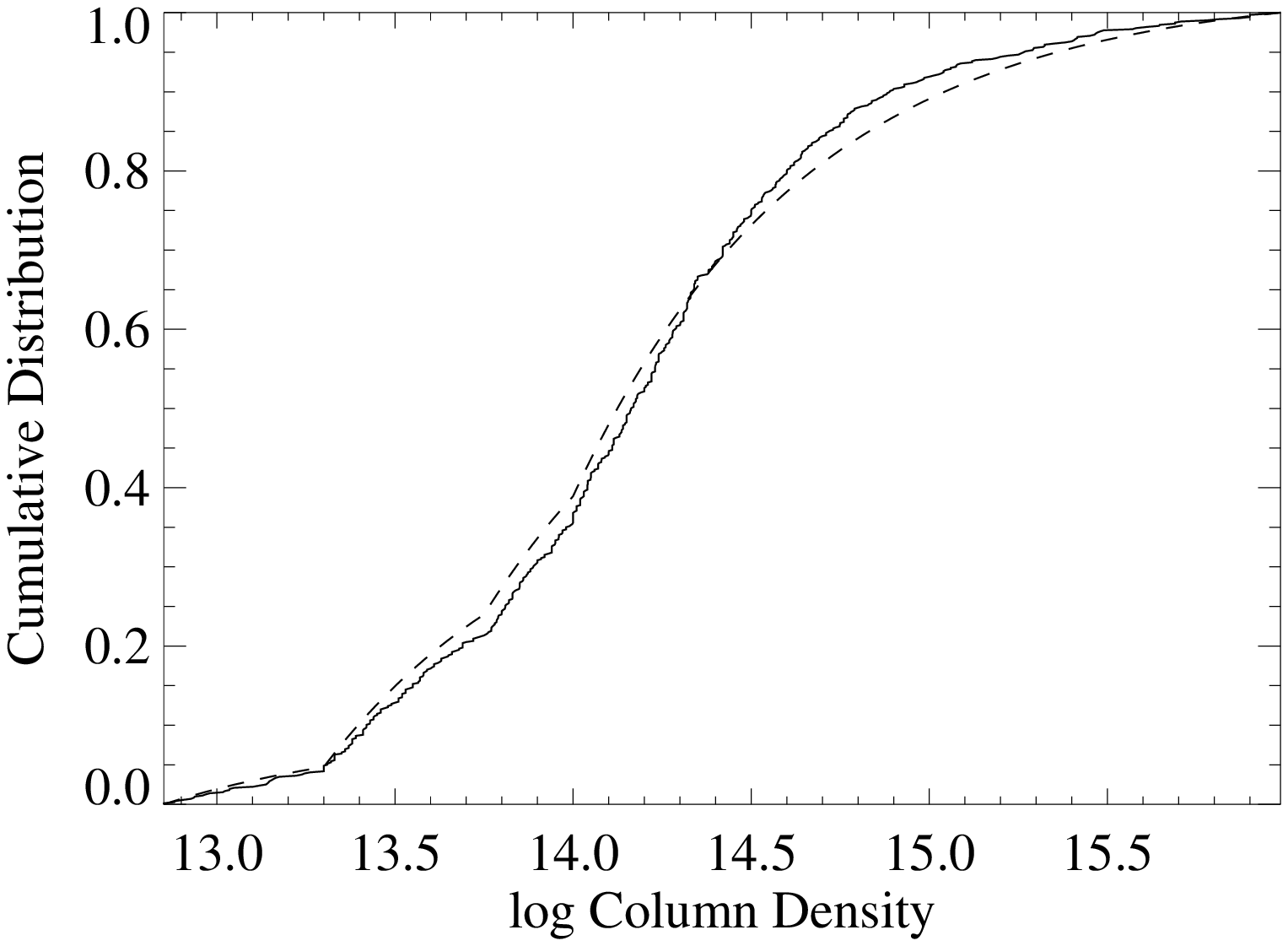}
\hfill
\epsfxsize=8.5cm
\epsfbox{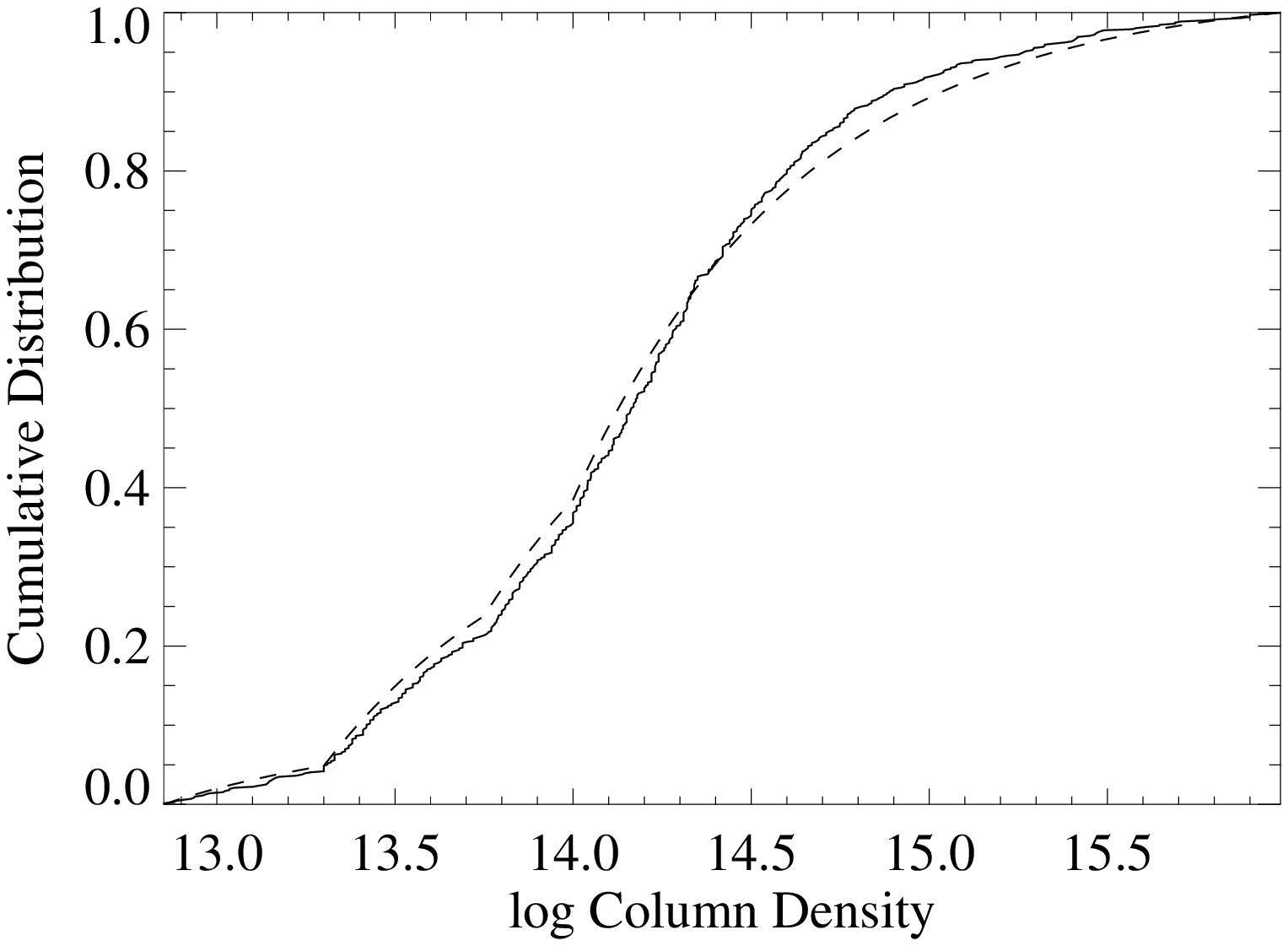}
}
\epsfverbosetrue
\caption{
The cumulative data (solid line) and model (dashed), integrating over
$z$, for the lines in the sample, plotted against column
density (\col).  The model on the right includes the proximity effect.}
\label{fig:cum1}
\end{figure*}

\begin{figure*}
\hbox{
\epsfxsize=8.5cm
\epsfbox{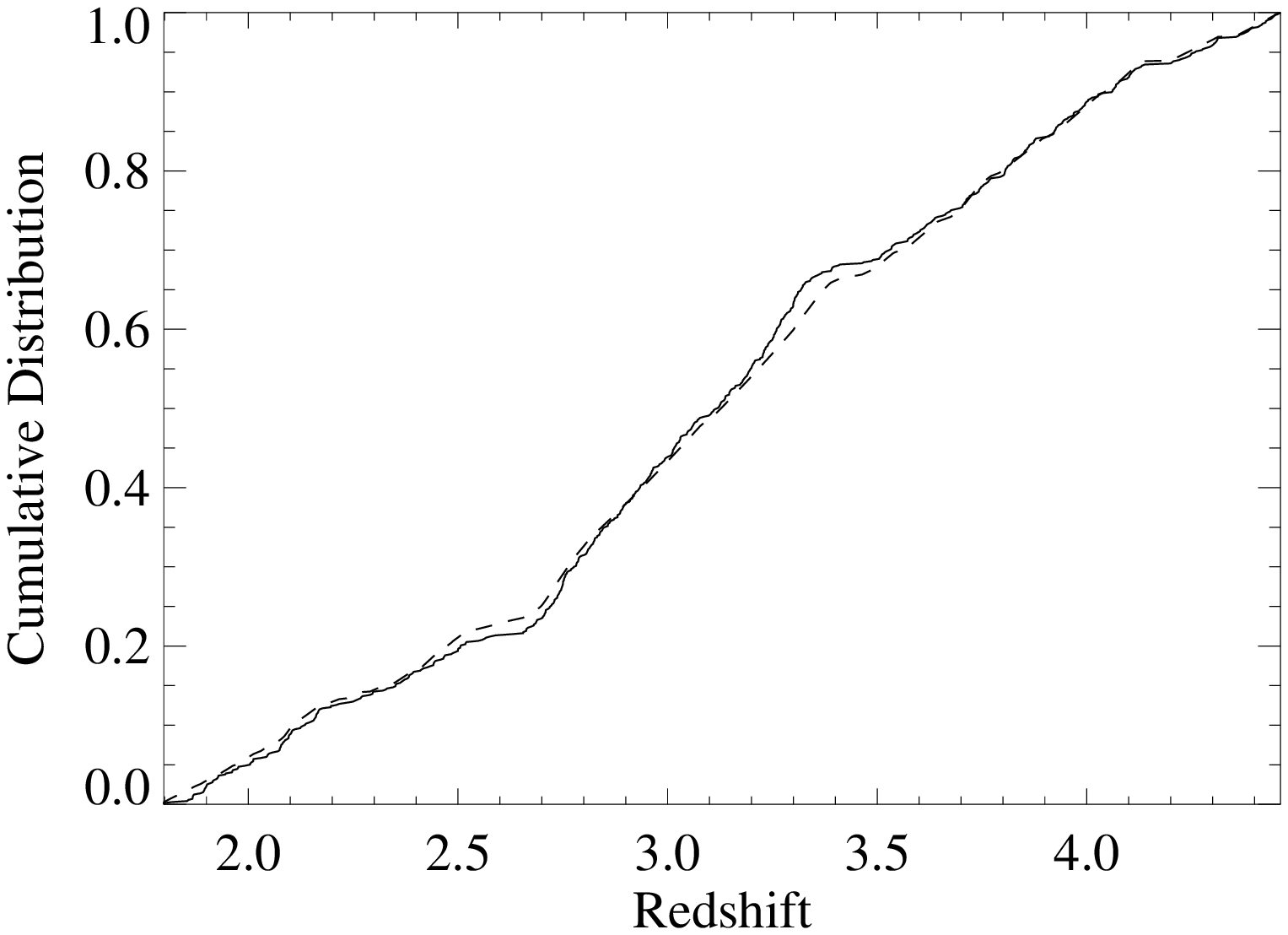}
\hfill
\epsfxsize=8.5cm
\epsfbox{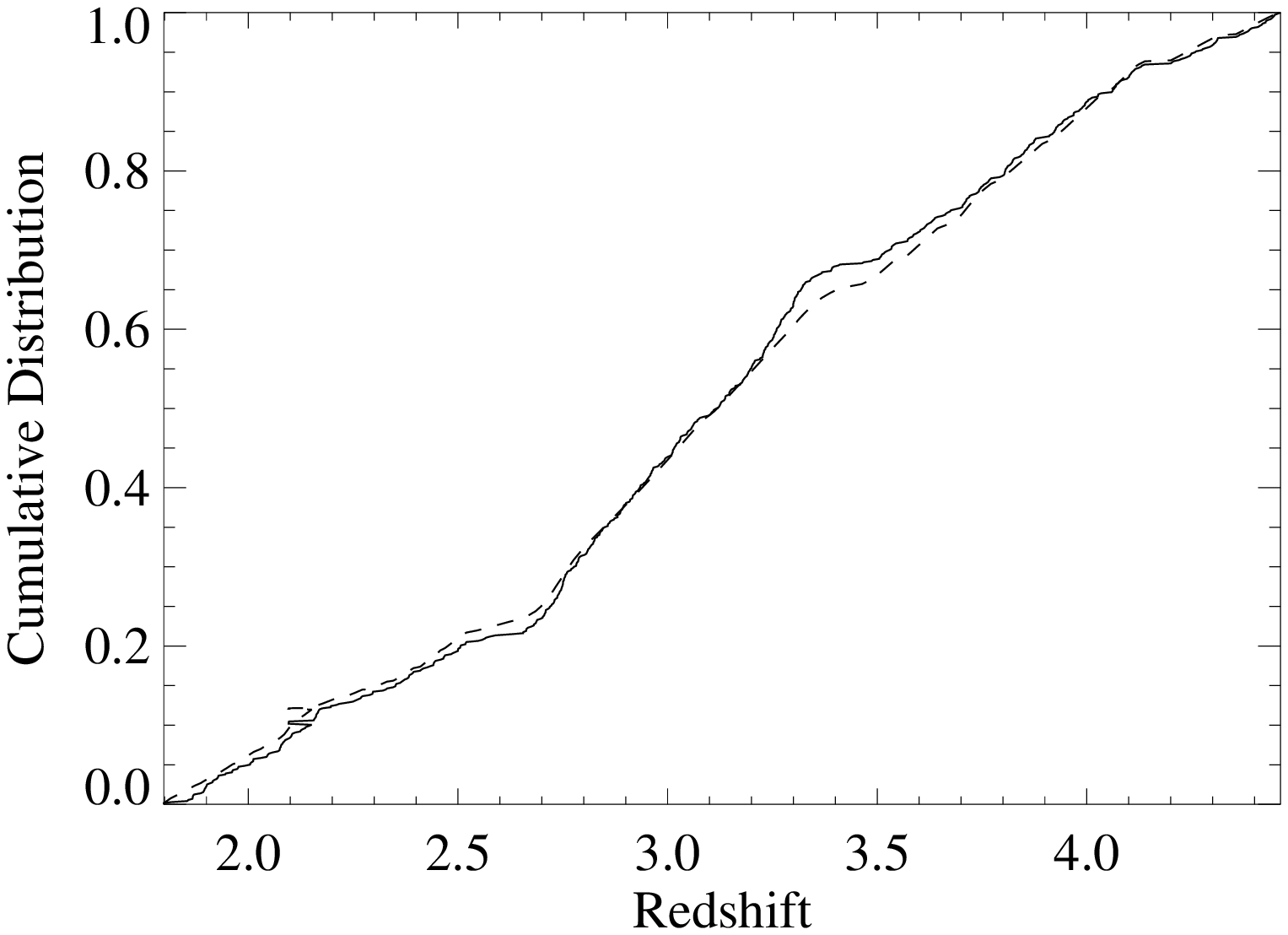}
}
\epsfverbosetrue
\caption{
The cumulative data (solid line) and model (dashed), integrating over
$N$, for the lines in the sample, plotted against redshift.
The model on the right includes the proximity effect.}
\label{fig:cum2}
\end{figure*}

Figures \ref{fig:cum1} and \ref{fig:cum2} show the cumulative data and
model for each variable using two models: one includes the proximity
effect (model {\bf B}), one does not (model {\bf A}). The
probabilities of the associated K--S statistics are given in table
\ref{tab:ks}.  For the column density plots the worst discrepancy
between model and data occurs at $\log(N)=14.79$.  The model with the
proximity effect (to the right) has slightly more high column density
clouds, as would be expected, although this is difficult to see in the
figures (note that the dashed line --- the model --- is the curve that
has changed).  In the redshift plots the difference between the two
models is more apparent because the changes are confined to a few
redshifts, near the quasars, rather than, as in the previous figures,
spread across a wide range of column densities.  The apparent
difference between model and data is larger for the model that
includes the proximity effect (on the right of figure~\ref{fig:cum2}).
However, this is an optical illusion as the eye tends to measure the
vertical difference between horizontal, rather than diagonal, lines.
In fact the largest discrepancy in the left figure is at $z=3.323$,
shifting to $z=3.330$ when the proximity effect is included.  It is
difficult to assess the importance of individual objects in cumulative
plots, but the main difference in the redshift figure occurs near the
redshift of Q0014+813.  However, since this is also the case without
the proximity effect (the left--hand figure) it does not seem to be
connected to the unusually large flux correction for this object
(section~\ref{sec:data}).

In both cases --- with and without the proximity effect --- the model
fits the data reasonably well.  It is not surprising that including
the proximity effect only increases the acceptability of the fit
slightly, as the test is dominated by the majority of lines which are
not influenced by the quasar.  The likelihood ratio test that we use
in section \ref{sec:disevid} is a more powerful method for comparing
two models, but can only be used if the models are already a
reasonable fit (as shown here).

%
%

\subsection{Sources of Error}
\label{sec:erress}

There are two sources of stochastic uncertainty in the values of
estimated parameters: the finite number of observations and the error
associated with each observation (column densities, redshifts, quasar
fluxes, etc.).

The first source of variation --- the limited information available
from a finite number of observations --- can be assessed by examining
the distribution of the posterior probabilities for each parameter.
This is described in the following section.

The second source of variation --- the errors associated with each
measurement --- can be assessed by repeating the analysis for
simulated sets of data.  In theory these errors could have been
included in the model and their contribution would have been apparent
in the posterior distribution.  In practice there was insufficient
information or computer time to make a detailed model of the error
distribution.

Instead, ten different sets of line--lists were created.  Each was
based on the original, with each new value, $X$, calculated from the
observed value $x$ and error estimate $\sigma_X$: \begin{equation} X =
x + a \sigma_X\ ,\end{equation} where $a$ was selected at random from
a (approximate) normal distribution with zero mean and unit variance.
The redshift (standard error 0.003) and luminosity (standard error 0.2
magnitudes) of each background quasar were also changed.  For
Q0420--388 the redshift error was increased to 0.1 and, for
Q1448--232, the magnitude error was increased to 0.6 magnitudes.  The
model was fitted to each data set and the most likely values of the
parameters recorded.  A Gaussian was fitted to the distribution of
values.  In some cases (eg.\ figure~\ref{fig:alphas_d}) a Gaussian
curve may not be the best way to describe the distribution of
measurements.  However, since the error in the parameters is dominated
by the small number of data points, rather than the observational
errors, using a different curve will make little difference to the
final results.

Since the two sources of stochastic error are not expected to be
correlated they can be combined to give the final distribution of the
parameters.  The Gaussian fitted to the variation from measurement
errors is convolved with the posterior distribution of the variable.
The final, normalized distribution is then a good approximation to the
actual distribution of values expected.

This procedure is shown in figures \ref{fig:beta_gamma_d}\ to
\ref{fig:alphas_d}.  For each parameter in the model the `raw'
posterior distribution is plotted (thin line and points).  The
distribution of values from the synthetic data is shown as a dashed
histogram and the fitted Gaussian is a thin line.  The final
distribution, after convolution, is the heavy line.  In general the
uncertainties due to a finite data set are the main source of error.

%
%

\subsection{Error Estimates from Posterior Probabilities\label{sec:postprob}}

\newcommand{\yb}{{\bf y}}
\newcommand{\tb}{{\bf\theta}} 
\newcommand{\rb}{{\bf R_\nu}} 
\newcommand{\bn}{{b_\nu}} 

If $p(\yb|\tb)$ is the likelihood of the observations ($\yb$), given
the model (with parameters $\tb$), then we need an expression for the
posterior probability of a `parameter of interest', $\eta$.  This
might be one of the model parameters, or some function of the
parameters (such as the background flux at a certain redshift):
\begin{equation}\eta = g(\tb)\ .\end{equation}

For example, the value of \jflux\ at a particular redshift for models
{\bf C} to {\bf E} in section~\ref{sec:popden} is a linear function of
several parameters (two or more of $J_{3.25}, J_{z_B}, \alpha_1$, and
$\alpha_2$).  To calculate how likely a particular flux is the
probabilities of all the possible combinations of parameter values
consistent with that value must be considered: it is necessary to
integrate over all possible values of $\beta$ and $\gamma^\prime$, and
all values of $J_{z_B}, \alpha_1$, etc. which are consistent with
\jflux(z) having that value.

In other words, to find the posterior distribution of $\eta$,
$\pi(\eta|\tb)$, we must marginalise the remaining model parameters:
\begin{equation}\pi(\eta|\tb)=\lim_{\gamma \rightarrow 0}
\frac{1}{\gamma} \int_D \pi(\tb|\yb)\,d\tb\ ,\end{equation} where D is
the region of parameter space for which $\eta \leq g(\tb) \leq \eta +
\gamma$ and $\pi(\tb|\yb) \propto \pi(\tb) p(\yb|\tb)$, the posterior
density of $\tb$ with prior $\pi(\tb)$.

A uniform prior is used here for all parameters (equivalent to normal
maximum likelihood analysis).  Explicitly, power law exponents and the
logarithm of the flux have prior distributions which are uniform over
$[-\infty,+\infty]$.

Doing the multi--dimensional integral described above would require a
large (prohibitive) amount of computer time.  However, the
log--likelihood can be approximated by a second order series expansion
in $\tb$.  This is equivalent to assuming that the other parameters
are distributed as a multivariate normal distribution, and the result
can then be calculated analytically.  Such a procedure is shown, by
Leonard, Hsu \& Tsui (1989), to give the following procedure when $g(\tb)$ is a
linear function of $\tb$: \begin{equation}\bar{\pi}(\eta|\yb) \propto
\frac{\pi_M(\eta|\yb)}{|\rb|^{1/2}(\bn^T\rb^{-1}\bn)^{1/2}}\
,\end{equation} where \begin{eqnarray} \pi_M(\eta|\yb) & = &
\sup_{\tb:g(\tb)=\eta} \pi(\tb|\yb)\\&=&\pi(\tb_\eta|\yb)\ ,\\ \bn & =
& \left.\frac{\partial g(\tb)}{\partial \tb}\right|_{\tb=\tb_\eta}\
,\\ \rb & = & \left.\frac{\partial^2 \ln
\pi(\tb|\yb)}{\partial(\tb\tb^T)} \right|_{\tb=\tb_\eta}\
. \end{eqnarray} The likelihood is maximised with the constraint that
$g(\tb)$ has a particular value.  $\rb$ is the Hessian matrix used in
the fitting routine (Press et~al. 1992) and $\bn$ is known (when $\eta$
is the average of the first two of three parameters, for example, $\bn
= 0.5,0.5,0$).

This quickens the calculation enormously.  To estimate the posterior
distribution for, say, \jflux, it is only necessary to choose a series
of values and, at each point, find the best fit consistent with that
value.  The Hessian matrix, which is returned by many fitting
routines, can then be used --- following the formulae above --- to
calculate an approximation to the integral, giving a value
proportional to the probability at that point.  Once this has been
repeated for a range of different values of \jflux\ the resulting
probability distribution can be normalised to give an integral of one.

Note that this procedure is only suitable when $g(\tb)$ is a linear
function of $\tb$ --- Leonard, Hsu \& Tsui (1989) give the expressions needed for
more complex parameters.

%
%

\section{Results}
\label{sec:results}

A summary of the results for the different models is given in table
\ref{tab:fpars}.  The models are:

\begin{description}

\item[{\bf A}] --- No Proximity Effect.  The population model
described in section \ref{sec:model}, but without the proximity
effect.

\item[{\bf B}] --- Constant Background.  The population model
described in section \ref{sec:model} with a constant ionising
background.

\item[{\bf C}] --- Power Law Background.  The population model
described in section \ref{sec:model} with an ionising background which
varies as a power law with redshift

\item[{\bf D}] --- Broken Power Law Background.  The population model
described in section \ref{sec:model} with an ionising background whose
power law exponent changes at $z_B=3.25$.

\item[{\bf E}] --- Correction for Extinction in Damped Systems.  As
{\bf D}, but with a correction for absorption in known damped
absorption systems (section \ref{sec:dampcor}).

\end{description}

In this paper we assume $q_0$ = 0.5 and $H_0$ = 100~\kms/Mpc.

%
%

\subsection{Population Distribution}
\label{sec:resmpars}

\begin{table*}
\begin{tabular}{crlrlrlrlrlrc}
Model & \multispan2{\hfill$\beta$\hfill} & \multispan2{\hfill$\gamma$\hfill} & \multispan2{\hfill $J_{z_B}$\hfill} & \multispan2{\hfill $\alpha_1$\hfill} & \multispan2{\hfill $\alpha_2$\hfill} & \multispan1{\hfill $z_B$\hfill} & -2 log--likelihood\\
 {\bf A} &  1.66 &  $\pm0.03$ &  2.7 &  $\pm0.3$ & \multicolumn{7}{c}{No background} & 60086.2 \\
 {\bf B} &  1.67 &  $\pm0.03$ &  2.9 &  $\pm0.3$ & $-21.0$ & $\pm0.2$ & & & & & & 60052.8 \\
 {\bf C} &  1.67 &  $\pm0.03$ &  3.0 &  $\pm0.3$ & $-21.0$ & $\pm0.2$ & $-1$ & $\pm3$ & & & & 60052.6 \\
 {\bf D} &  1.67 &  $\pm0.04$ &  3.0 &  $\pm0.3$ & $-20.9$ & $\pm0.3$ & 0 & $+5,-6$ & $-2$ & $\pm4$ & 3.25 & 60052.4 \\
 {\bf E} &  1.67 &  $\pm0.03$ &  3.0 &  $\pm0.3$ & $-20.9$ & $+0.3,-0.2$ & 0 & $+5,-6$ & $-2$ & $+7,-4$ & 3.25 & 60051.4 \\
\end{tabular}
\caption{
The best--fit parameters and expected errors for the models.}
\label{tab:fpars}
\end{table*}

The maximum likelihood `best--fit' values for the parameters are given
in table \ref{tab:fpars}.  The quoted errors are the differences (a
single value if the distribution is symmetric) at which the
probability falls by the factor $1/\sqrt{e}$.  This is equivalent to a
`$1\sigma$ error' for parameters with normal error distributions.

The observed evolution of the number of clouds per unit redshift is
described in the standard notation found in the literature
\begin{equation}dN/dz = A_0 (1+z)^{\gamma}\ .\end{equation} The
variable used in the maximum likelihood fits here, $\gamma^\prime$,
excludes variations expected from purely cosmological variations and
is related to $\gamma$ by: \begin{equation} \gamma = \left\{
\begin{array}{ll}\gamma^\prime + 1 & \mbox{ if $q_0 = 0$} \\
\gamma^\prime + \frac{1}{2} & \mbox{ if $q_0 = 0.5$ \ .}\end{array}
\right.\end{equation}

\begin{figure*}
\hbox{
\epsfxsize=8.5cm
\epsfbox{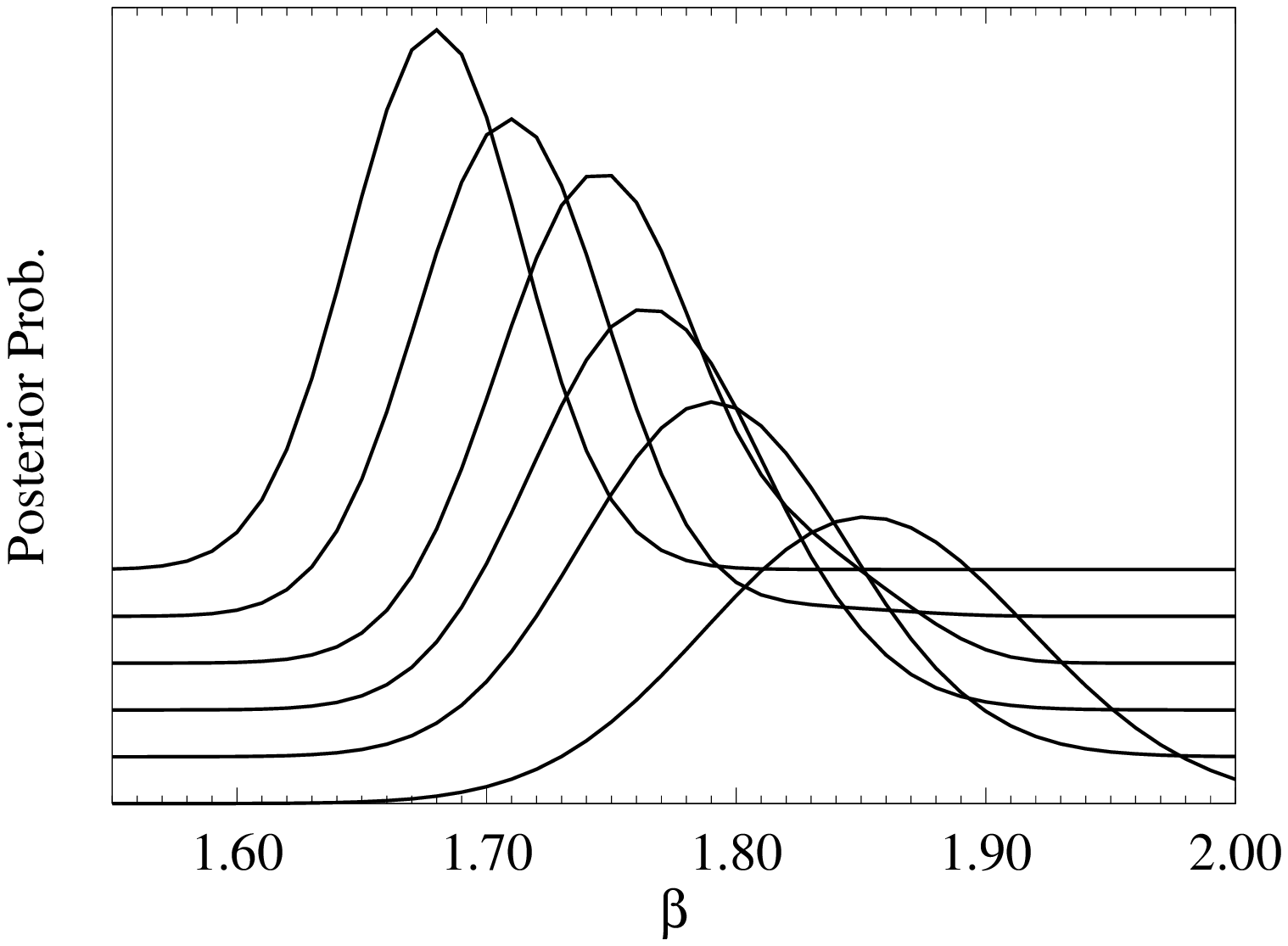}
\hfill
\epsfxsize=8.5cm
\epsfbox{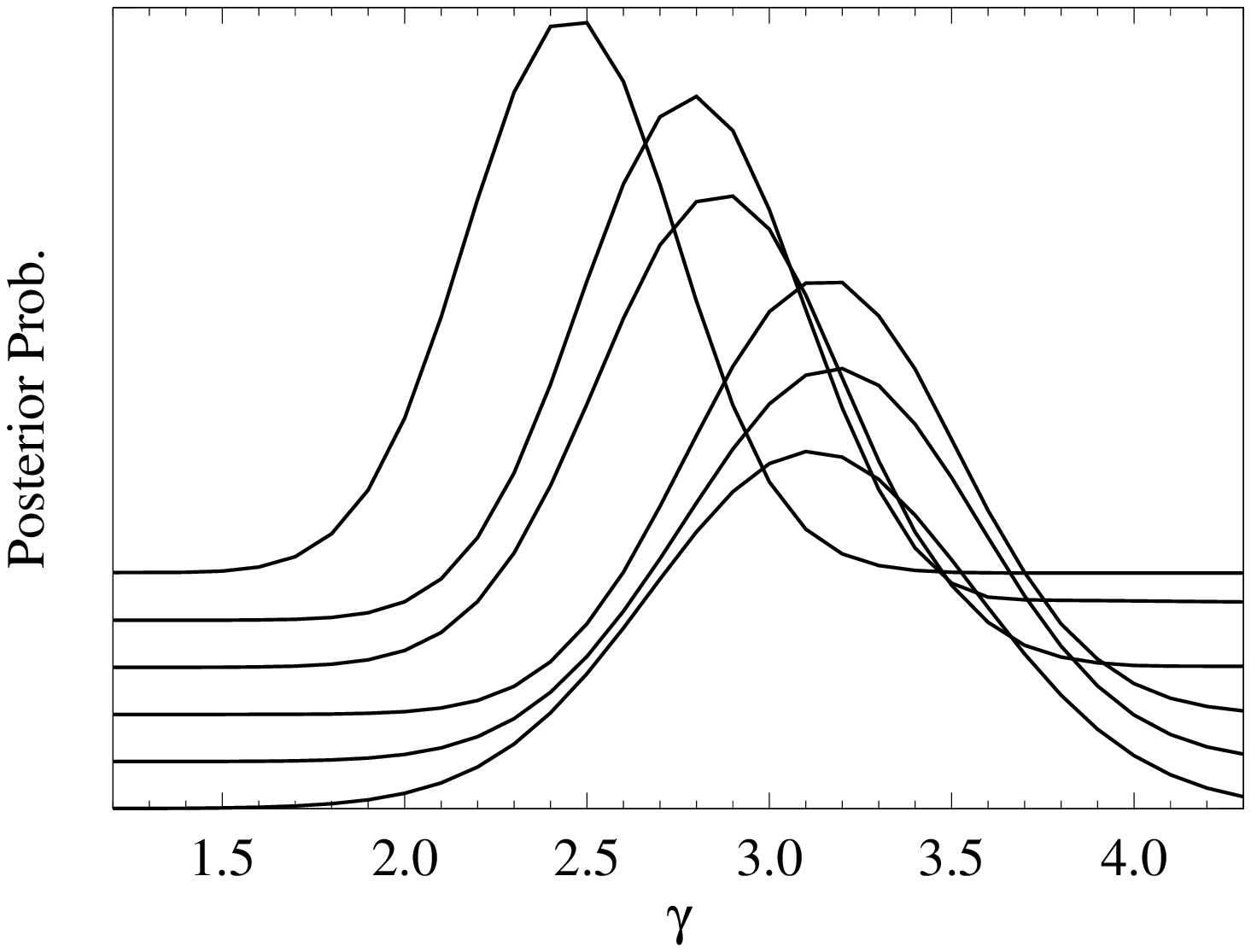}
}
\epsfverbosetrue
\caption{The expected probability distribution of the population parameters for model {\bf D}.  The top curve is for all data, each lower curve is for data remaining when the column density completeness limits are progressively increased by $\Delta\col=0.1$.}
\label{fig:var_d}
\end{figure*}

Figure \ref{fig:var_d}\ shows the variation in population parameters
for model {\bf D} as the completeness limits are increased in steps of
$\Delta\col=0.1$.  The number of clouds decreases from 844 to 425
(when the completeness levels have been increased by
$\Delta\col=0.5$).

\begin{figure*}
\hbox{
\epsfxsize=8.5cm
\epsfbox{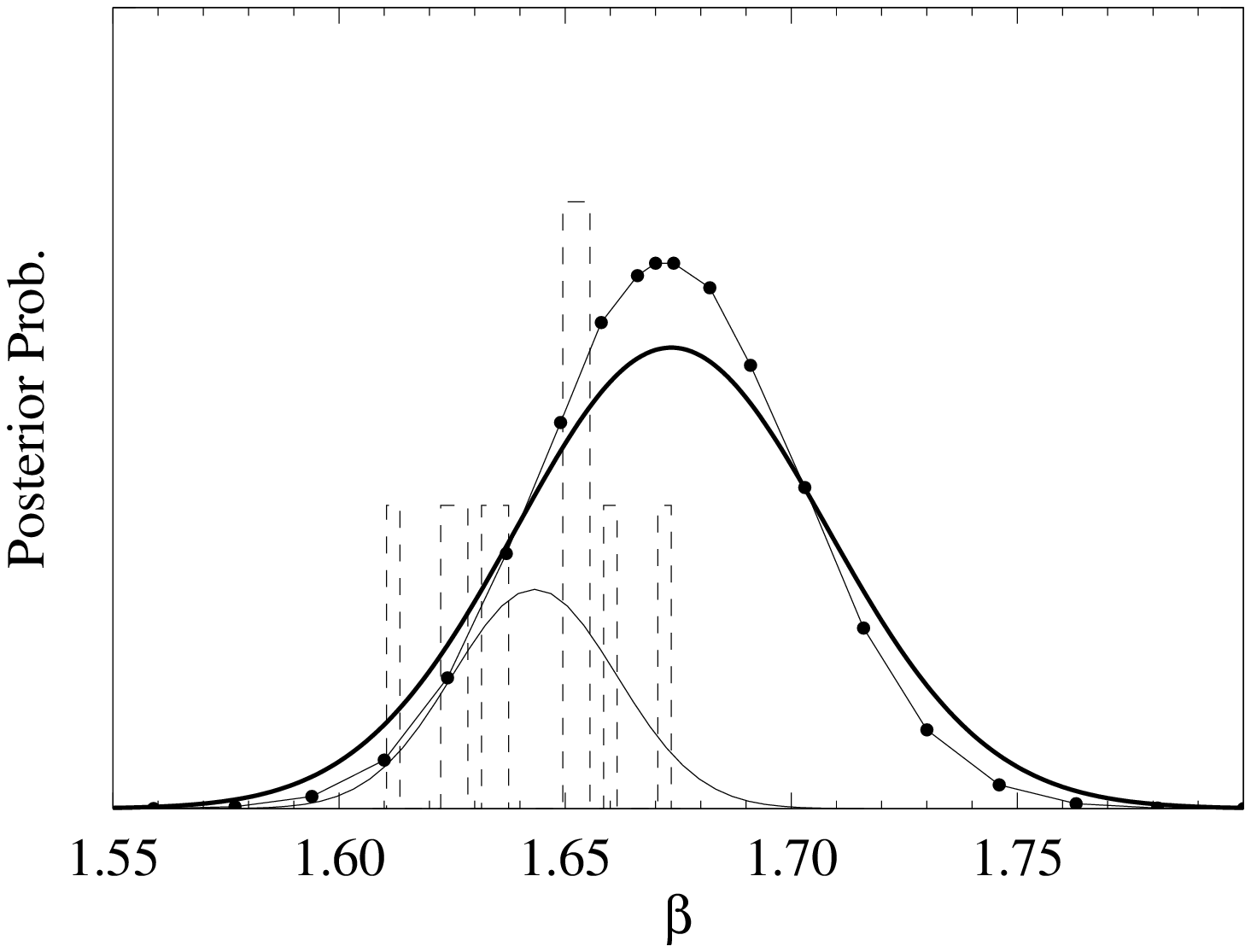}
\hfill
\epsfxsize=8.5cm
\epsfbox{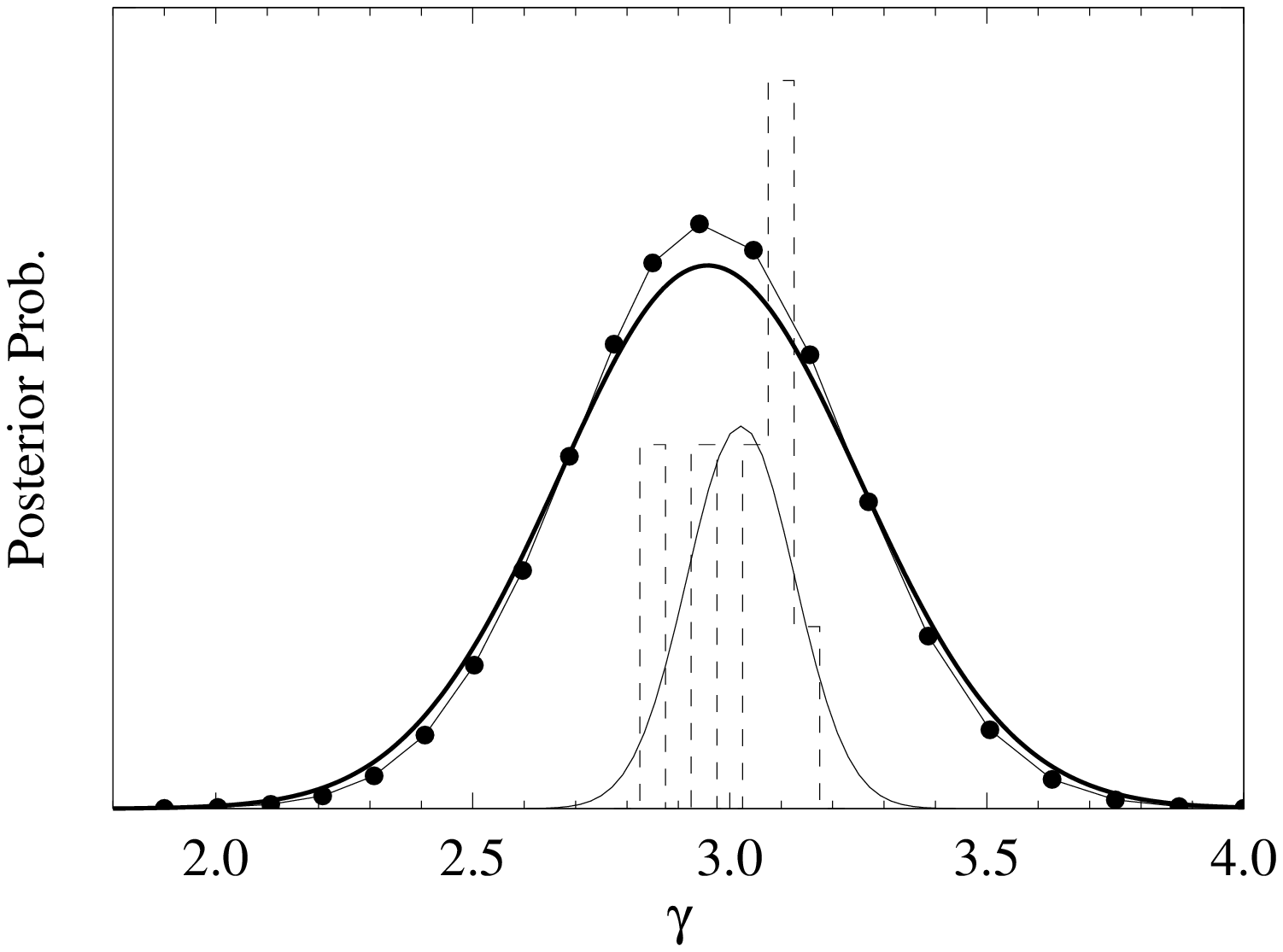}
}
\epsfverbosetrue
\caption{The expected probability distribution of the model parameters $\beta$ and $\gamma$ (heavy line) for model {\bf D}.  The dashed histogram and Gaussian (thin line) show how the measured value varies for different sets of data.  The dash-dot line shows the uncertainty in the parameter because the data are limited.  These are combined to give the final distribution (bold).  See section \ref{sec:erress} for mode details.}
\label{fig:beta_gamma_d}
\end{figure*}

\begin{figure*}
\hbox{
\epsfxsize=8.5cm
\epsfbox{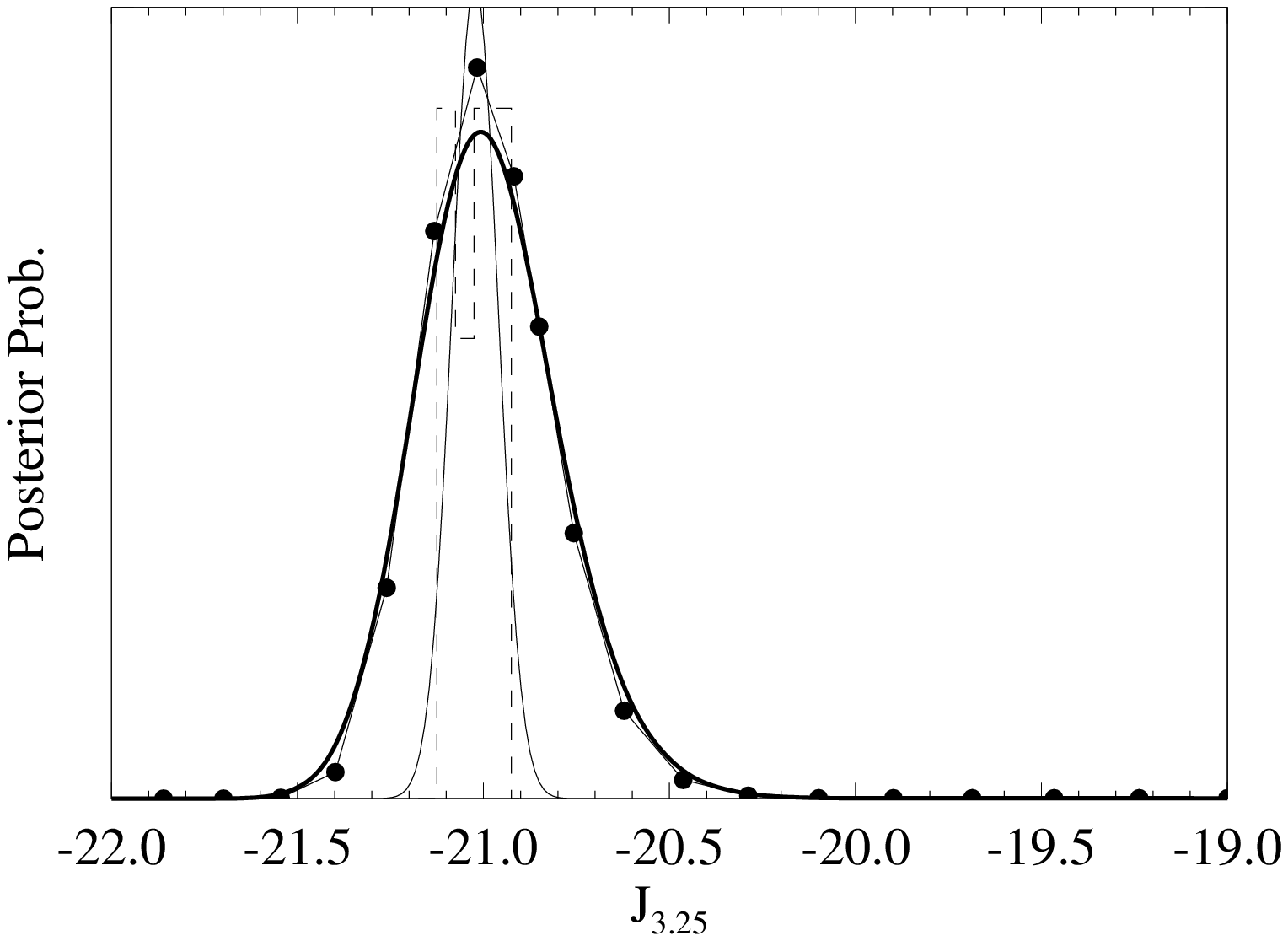}
\hfill
\epsfxsize=8.5cm
\epsfbox{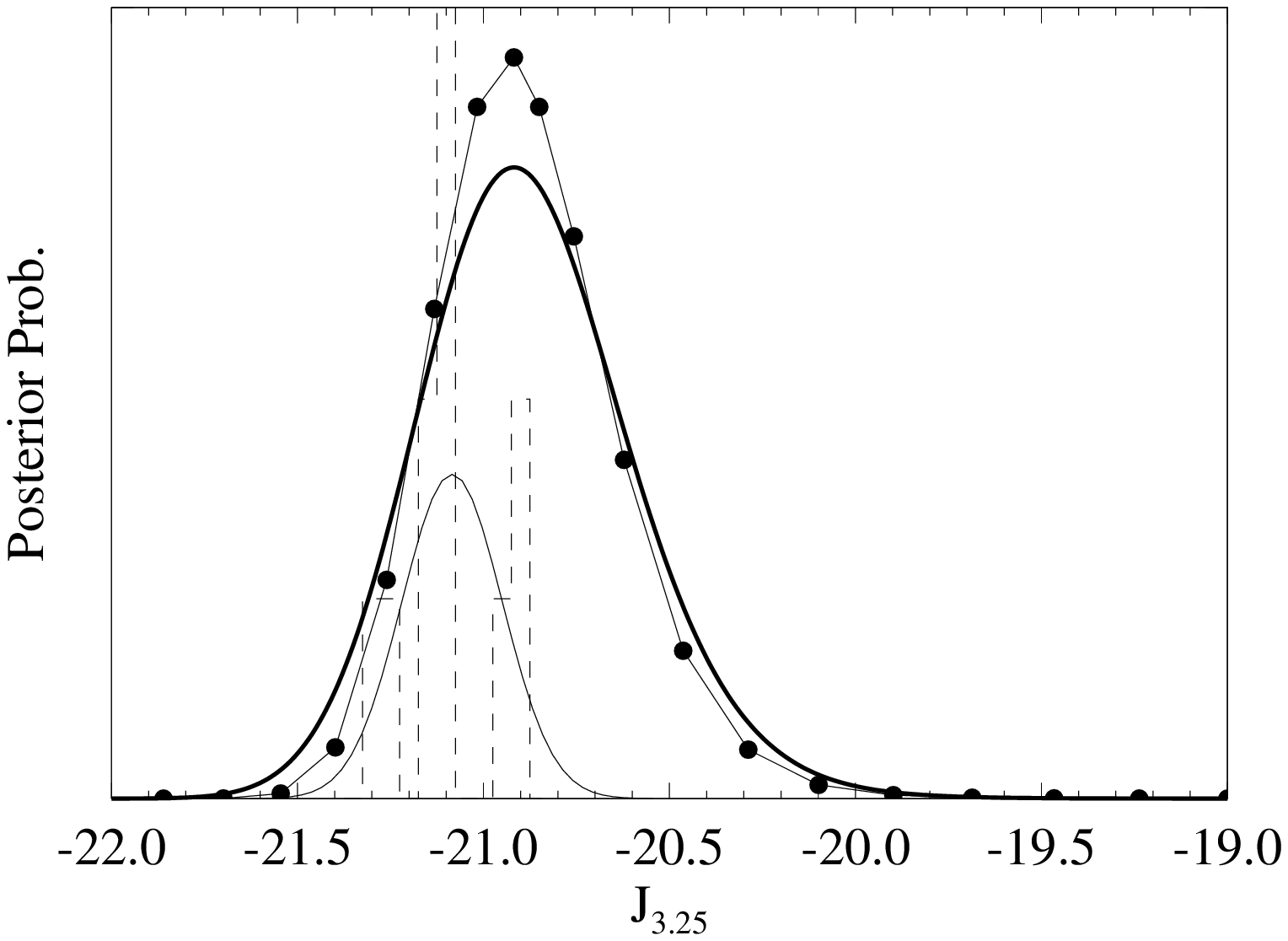}
}
\epsfverbosetrue
\caption{The expected probability distribution of the log background flux at $z=3.25$ (heavy line) for models {\bf B} (left) and {\bf D}.  The uncertainty from the small number of lines near the quasar (line with points) is significantly larger than that from uncertainties in column densities or quasar properties (thin curve).  See section \ref{sec:erress} for a full description of the plot.}
\label{fig:nu1_d}
\end{figure*}

\begin{figure*}
\hbox{
\epsfxsize=8.5cm
\epsfbox{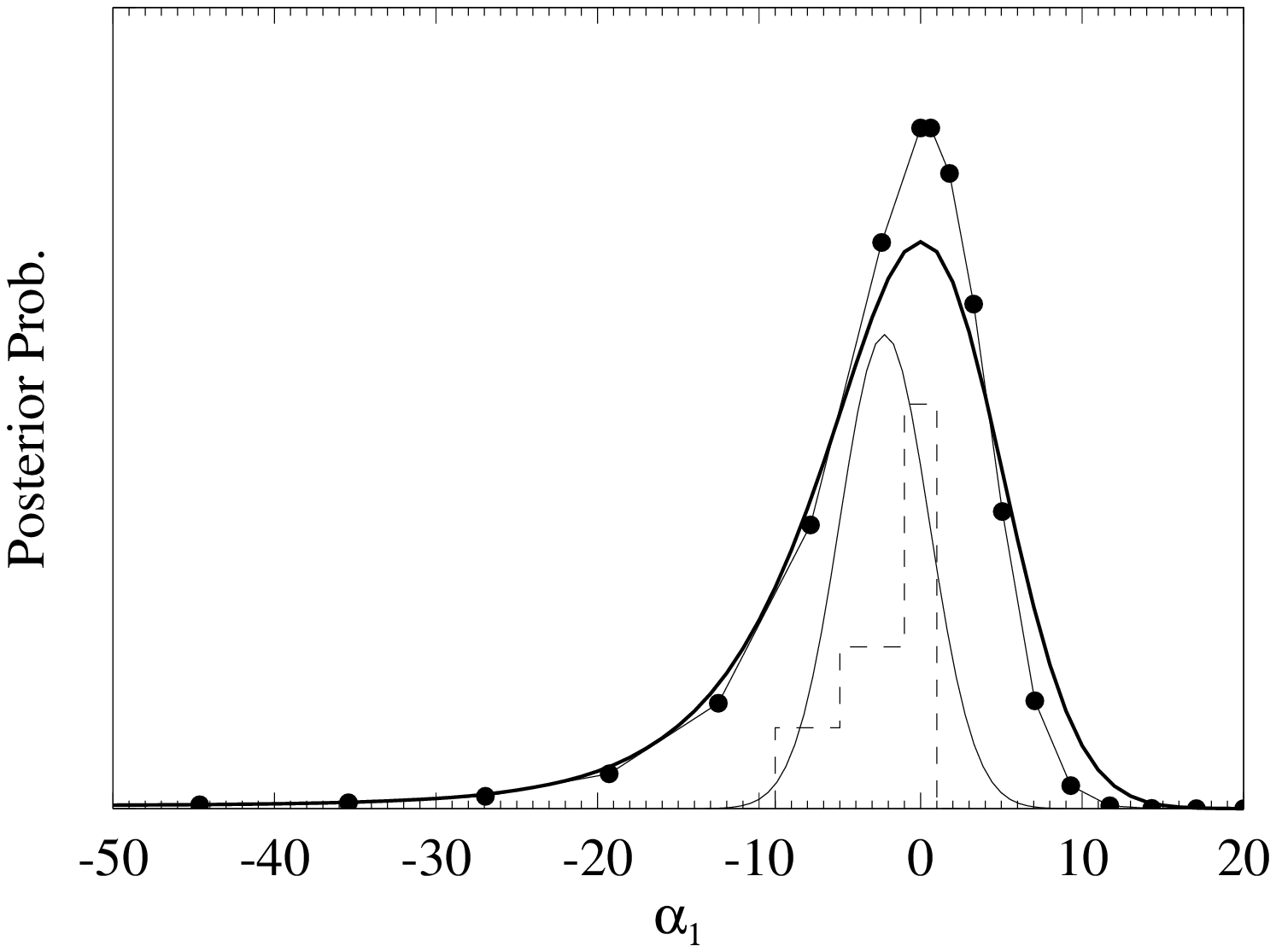}
\hfill
\epsfxsize=8.5cm
\epsfbox{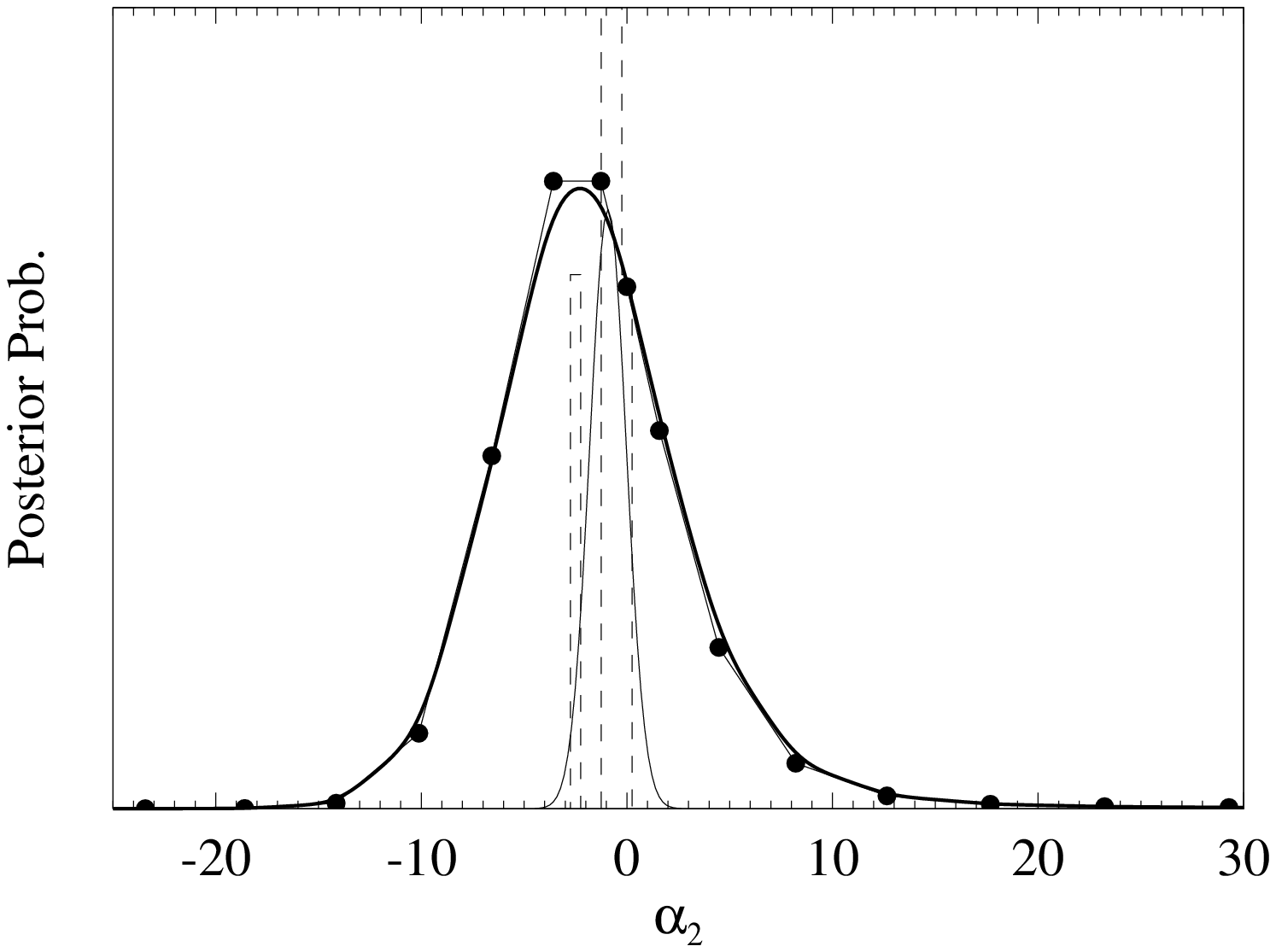}
}
\epsfverbosetrue
\caption{The expected probability distribution of the model parameters $\alpha_1$ and $\alpha_2$ (heavy line) for model {\bf D}.  See section \ref{sec:erress} for a full description of the plot.}
\label{fig:alphas_d}
\end{figure*}

%
%

\subsection{Ionising Background}
\label{sec:resbackg}

Values of the ionising flux parameters are show in table
\ref{tab:fpars}.  The expected probability distributions for models
{\bf B} and {\bf D} are shown in figures \ref{fig:nu1_d} and
\ref{fig:alphas_d}.  The background flux relation is described in
section \ref{sec:popden}.

\begin{figure*}
\hbox{
\epsfxsize=8.5cm
\epsfbox{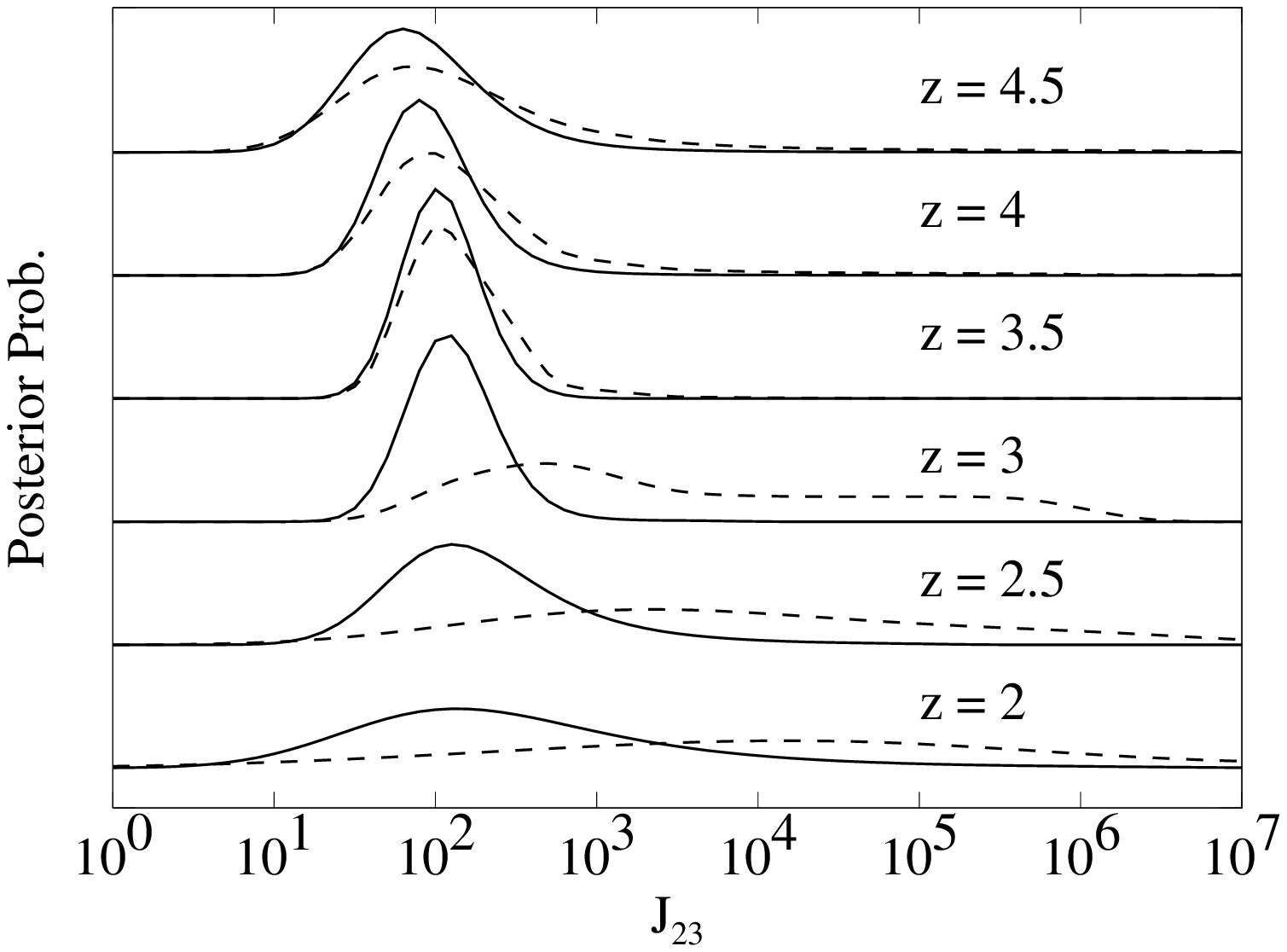}
\hfill
\epsfxsize=8.5cm
\epsfbox{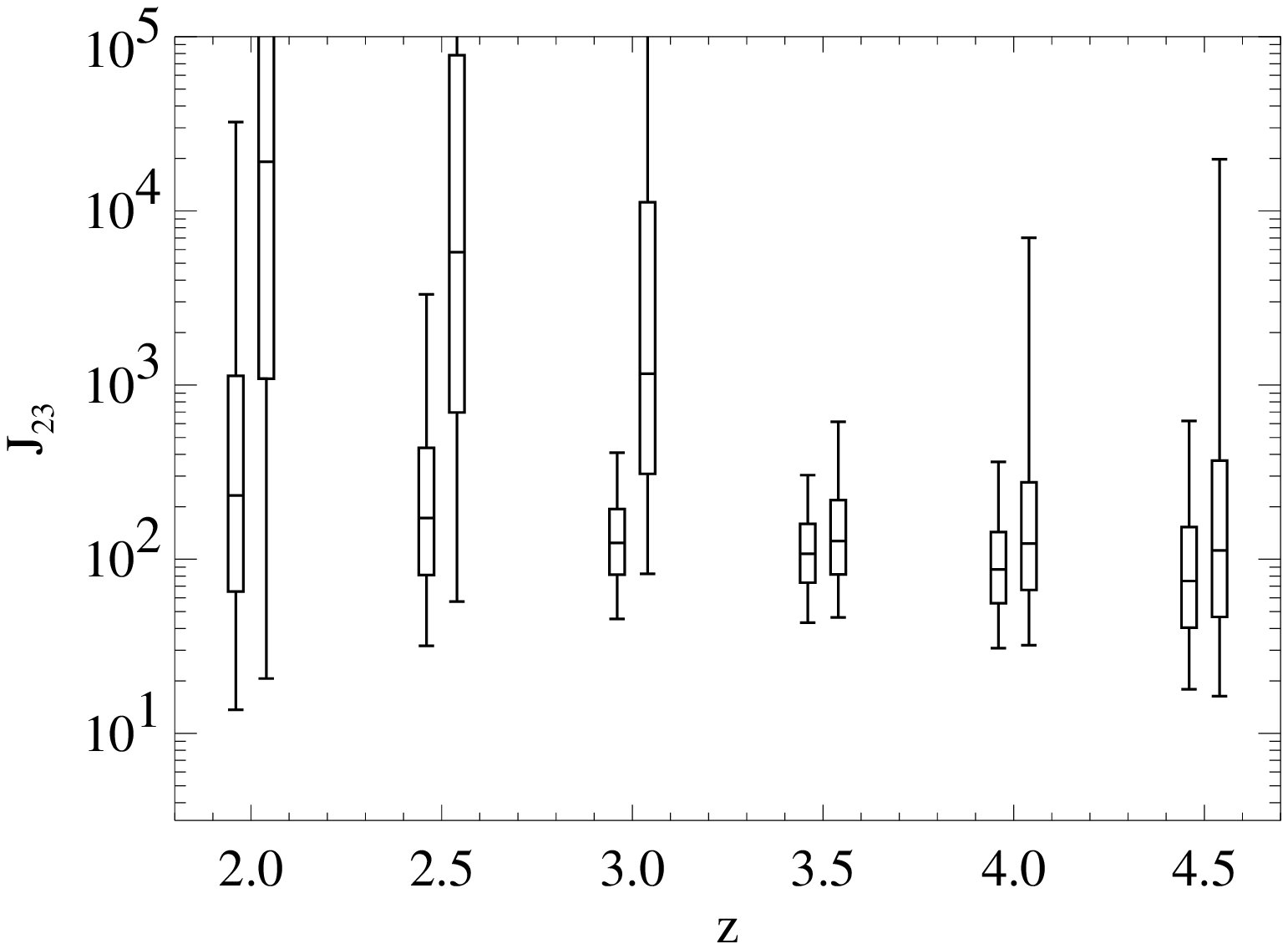}
}
\epsfverbosetrue
\caption{The expected probability distribution of the log background flux for model {\bf D}, comparing the results from the full data set with those obtained when the column density completeness limit is raised by $\Delta\log(N)=0.5$ (dashed line, left; larger boxes, right).  The box plots show median, quartiles, and 95\% limits.}
\label{fig:flux_both}
\end{figure*}

The variables used to describe the variation of the flux with redshift
are strongly correlated.  To illustrate the constraints more clearly
the marginalised posterior distribution (section \ref{sec:postprob})
of \jflux\ was calculated at a series of redshifts.  These are shown
(after convolution with the combination of Gaussians appropriate for
the uncertainties in the parameters from observational errors) for
model {\bf D} in figure \ref{fig:flux_both}.  The distribution at each
redshift is calculated independently.  This gives a conservative
representation since the marginalisation procedure assumes that
parameters can take all possible values consistent with the background
at that redshift (the probability that the flux can be low at a
certain redshift, for example, includes the possibility that it is
higher at other redshifts).  Figure \ref{fig:flux_both} also compares
the results from the full data set (solid lines and smaller boxes)
with those from the data set with column density completeness limits
raised by $\Delta\log(N)=0.5$ (the same data as the final curves in
figure \ref{fig:var_d}).

Table~\ref{tab:modeld} gives the most likely flux (at probability
$p_m$), an estimate of the `1$\sigma$ error' (where the probability
drops to $p_m/\sqrt{e}$), the median flux, the upper and lower
quartiles, and the 5\% and 95\% limits for model {\bf D} at the
redshifts shown in figure~\ref{fig:flux_both}.  It is difficult to
assess the uncertainty in these values.  In general the central
measurements are more reliable than the extreme limits.  The latter
are more uncertain for two reasons.  First, the distribution of
unlikely models is more likely to be affected by assumptions in
section~\ref{sec:postprob} on the normal distribution of secondary
parameters.  Second, the tails of the probability distribution are
very flat, making the flux value sensitive to numerical noise.
Extreme limits, therefore, should only be taken as a measure of the
relevant flux magnitude.  Most likely and median values are given to
the nearest integer to help others plot our results --- the actual
accuracy is probably lower.

\begin{table}
\begin{tabular}{crrrrrr}
&$z=2$&$z=2.5$&$z=3$&$z=3.5$&$z=4$&$z=4.5$\\
$p_m/\sqrt{e}$           &  30 &  50 &  60 &  60 &  40 &  30 \\
$p_m$                    & 137 & 129 & 118 & 103 &  80 &  63 \\
$p_m/\sqrt{e}$           &1000 & 400 & 220 & 180 & 160 & 170 \\
5\%                      &  10 &  30 &  50 &  40 &  30 &  20 \\
25\%                     &  70 &  80 &  80 &  70 &  60 &  40 \\
50\%                     & 232 & 172 & 124 & 108 &  87 &  75 \\
75\%                     &1000 & 400 & 200 & 160 & 100 & 200 \\
95\%                     &30000&3000 & 400 & 300 & 400 & 600 \\
\end{tabular}
\caption{
The fluxes (\jflux) corresponding to various posterior probabilities for model {\bf D}.  See the text for details on the expected errors in these values.}
\label{tab:modeld}
\end{table}

%
%

\section{Discussion}
\label{sec:discuss}

%
%

\subsection{Population Parameters}
\label{sec:dispop}

Parameter values for the different models are given in table
\ref{tab:fpars}.  They are generally consistent with other estimates
(Lu, Wolfe \& Turnshek 1991; Rauch et~al. 1993).  Including the proximity effect
increases $\gamma$ by $\sim0.2$.  Although not statistically
significant, the change is in the sense expected, since local
depletions at the higher redshift end of each data set are removed.

Figure \ref{fig:var_d}\ shows the change in population parameters as
the completeness limits for the observations are increased.  The most
likely values (curve peaks) of both $\beta$ and $\gamma$ increase as
weaker lines are excluded, although $\gamma$ decreases again for the
last sample.  

The value of $\beta$ found here ($\sim 1.7$) is significantly
different from that found by Press \& Rybicki (1993) ($\beta \sim 1.4$) using a
different technique (which is insensitive to Malmquist bias and line
blending).  The value of $\beta$ moves still further away as the
column density completeness limits are increased.  This is not
consistent with Malmquist bias, which would give a smaller change in
$\beta$ (section~\ref{sec:malm}), but could be a result of either line
blending or a population in which $\beta$ increases with column
density.  The latter explanation is also consistent with
Cristiani et~al. (1995) who found a break in the column density distribution
with $\beta$ = 1.10 for log($N$)$ < 14.0$, and $\beta$ = 1.80 above
this value.  Later work (Giallongo et~al. 1996, see section~\ref{sec:prevhi})
confirmed this.

Recent work by Hu et~al. (1995), however, using data with better
signal--to--noise and resolution, finds that the distribution of
column densities is described by a single power law ($\beta\sim 1.46$)
until $\col\sim 12.3$, when line blending in in their sample becomes
significant.  It might be possible that their sample is not
sufficiently large (66 lines with $\log(N)>14.5$, compared with 192
here) to detect a steeper distribution of high column density lines.

The change in $\gamma$ as completeness limits are raised may reflect
the decrease in line blending at higher column densities.  This
suggests that the value here is an over--estimate, although the shift
is within the 95\% confidence interval.  No estimate is significantly
different from the value of 2.46 found by Press \& Rybicki (1993) (again, using
a method less susceptible to blending problems).

%
%

\subsection{The Proximity Effect}

%
%

\subsubsection{Is the Proximity Effect Real?}
\label{sec:disevid}

The likelihood ratio statistic (equivalent to the `F test'), comparing
model {\bf A} with any other, indicates that the null hypothesis (that
the proximity effect, as described by the model here, should be
disregarded) can be rejected with a confidence exceeding 99.9\%.  Note
that this confirmation is based on the likelihood values in
table~\ref{tab:fpars}.  This test is much more powerful than the K--S
test (section~\ref{sec:finalq}) which was only used to see whether the
models were sufficiently good for the likelihood ratio test to be
used.

To reiterate: if model {\bf A} and model {\bf B} are taken as
competing descriptions of the population of \lya\ clouds, then the
likelihood ratio test, which allows for the extra degree of freedom
introduced, strongly favours a description which includes the
proximity effect.  The model without the proximity effect is firmly
rejected.  This does not imply that the interpretation of the effect
(ie.\ additional ionization by background radiation) is correct, but
it does indicate that the proximity effect, in the restricted,
statistical sense above, is `real' (cf. R\"{o}ser 1995).

If the assumptions behind this analysis are correct, in particular
that the proximity effect is due to the additional ionising flux from
the quasar, then the average value of the background is
100\elim{50}{30}~\jflux\ (model {\bf B}).

If a more flexible model for the background (two power laws) is used
the flux is consistent with a value of 120\elim{110}{50}~\jflux\
(model {\bf D} at $z=3.25$).

%
%

\subsubsection{Systematic Errors\label{sec:syserr}}

Five sources of systematic error are discussed here: Malmquist bias
and line blending; reddening by damped absorption systems; increased
clustering of clouds near quasars; the effect of gravitational
lensing.

The constraints on the background given here may be affected by
Malmquist bias and line blending (sections \ref{sec:malm} and
\ref{sec:dispop}).  The effects of line blending will be discussed
further in section~\ref{sec:gerry}, where a comparison with a
different procedure suggests that it may cause us to over--estimate
the flux (by perhaps $0.1$ dex).  Malmquist bias is more likely to
affect parameters sensitive to absolute column densities than those
which rely only on relative changes in the observed population.  So
while this may have an effect on $\beta$, it should have much less
influence on the inferred background value.

Attenuation by intervening damped absorption systems will lower the
apparent quasar flux and so give an estimate for the background which
is too low.  This is corrected in model {\bf E}, which includes
adjustments for the known damped systems (section \ref{sec:dampcor},
table~\ref{tab:damp}).  The change in the inferred background flux is
insignificant (figure \ref{fig:evoln}, table~\ref{tab:fpars}),
implying that the magnitude of the bias is less than 0.1 dex.

If quasars lie in regions of increased absorption line clustering
(Loeb \& Eisenstein 1995; Yurchenko, Lanzetta \& Webb 1995) then the background flux may be
overestimated by up to 0.5, or even 1, dex.

Gravitational lensing may change the apparent brightness of a quasar
--- in general the change can make the quasar appear either brighter
or fainter.  Absorption line observations are made towards the
brightest quasars known (to get good quality spectra).  Since there
are more faint quasars than bright ones this will preferentially
select objects which have been brightened by lensing (see the comments
on Malmquist bias in section~\ref{sec:malm}).  An artificially high
estimate of the quasar flux will cause us to over--estimate the
background.

Unfortunately, models which assess the magnitude of the increase in
quasar brightness are very sensitive to the model population of
lensing objects.  From Pei (1995) an upper limit consistent with
observations is an increase in flux of about 0.5 magnitudes,
corresponding to a background estimate which is too high by a factor
of 1.6 (0.2 dex).  The probable effect, however, could be much smaller
(Blandford \& Narayan 1992).

If bright quasars are more likely to be lensed we can make a
rudimentary measurement of the effect by splitting the data into two
separate samples.  When fitted with a constant background (model {\bf
B}) the result for the five brightest objects is indeed brighter than
that for the remaining six, by 0.1 dex.  The errors, however, are
larger (0.3 dex), making it impossible to draw any useful conclusions.

The effects of Malmquist bias, line blending and damped absorption
systems are unlikely to change the results here significantly.  Cloud
clustering and gravitational lensing could be more important --- in
each case the background would be over--estimated.  The magnitude of
these last two biases is not certain, but cloud clustering seems more
likely to be significant.

%
%

\subsubsection{Is there any Evidence for Evolution?}
\label{sec:noevoln}

More complex models allow the background flux to vary with redshift.
If the flux does evolve then these models should fit the data better.
However, there is no significant change in the fit when comparing the
likelihood of models {\bf C} to {\bf E} with that of {\bf B}.  Nor are
$\alpha_1$ or $\alpha_2$ significantly different from zero.  So there
is no significant evidence for a background which changes with
redshift.

The asymmetries in the wings of the posterior distributions of
$\alpha_1$ or $\alpha_2$ for model {\bf D} (figure \ref{fig:alphas_d})
are a result of the weak constraints on upper limits (see next
section).  The box plots in figure \ref{fig:flux_both} illustrate the
range of evolutions that are possible.

%
%

\subsubsection{Upper and Lower Limits\label{sec:lims}}

\begin{figure*}
\hbox{
\epsfxsize=8.5cm
\epsfbox{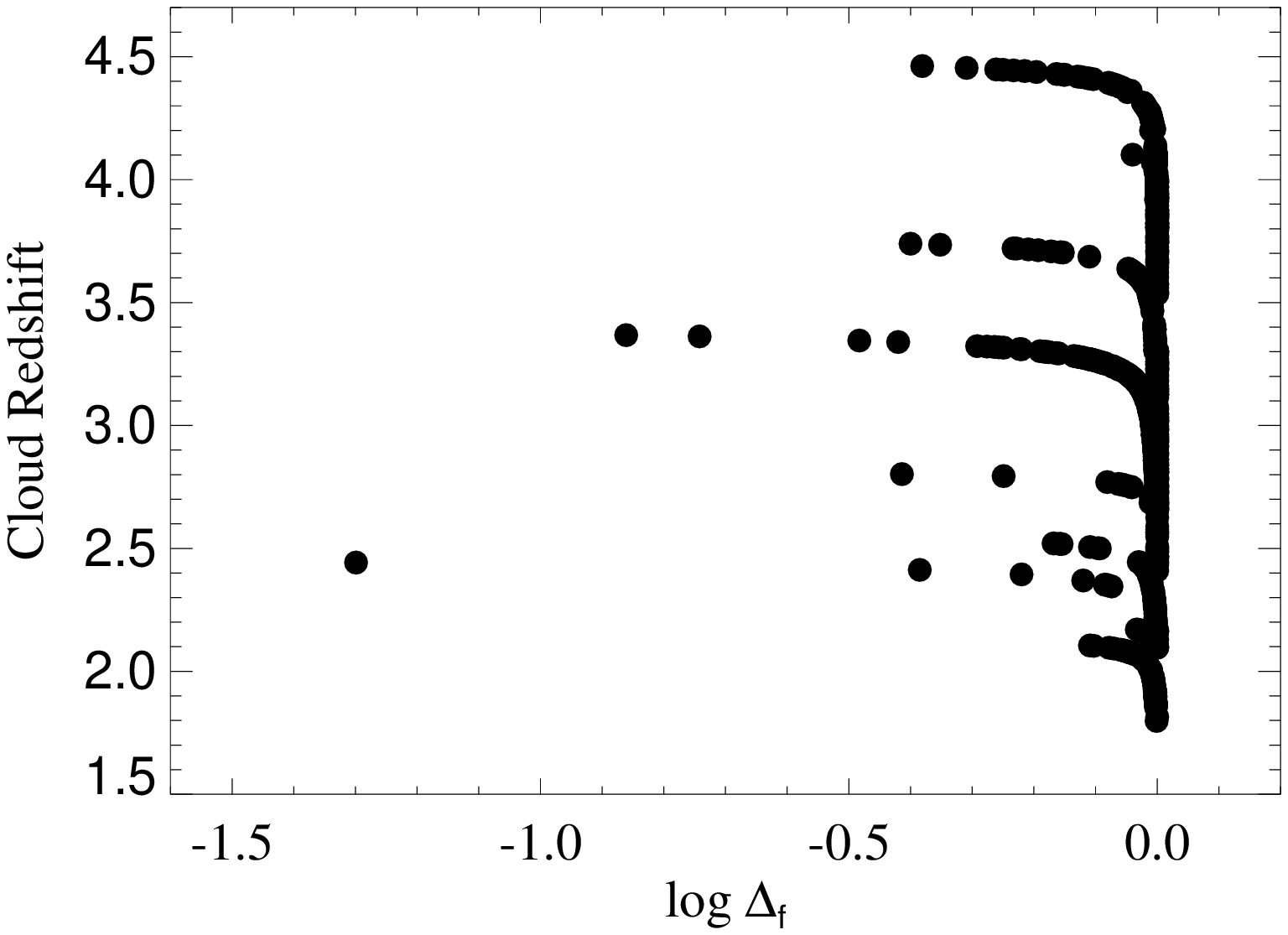}
\hfill
\epsfxsize=8.5cm
\epsfbox{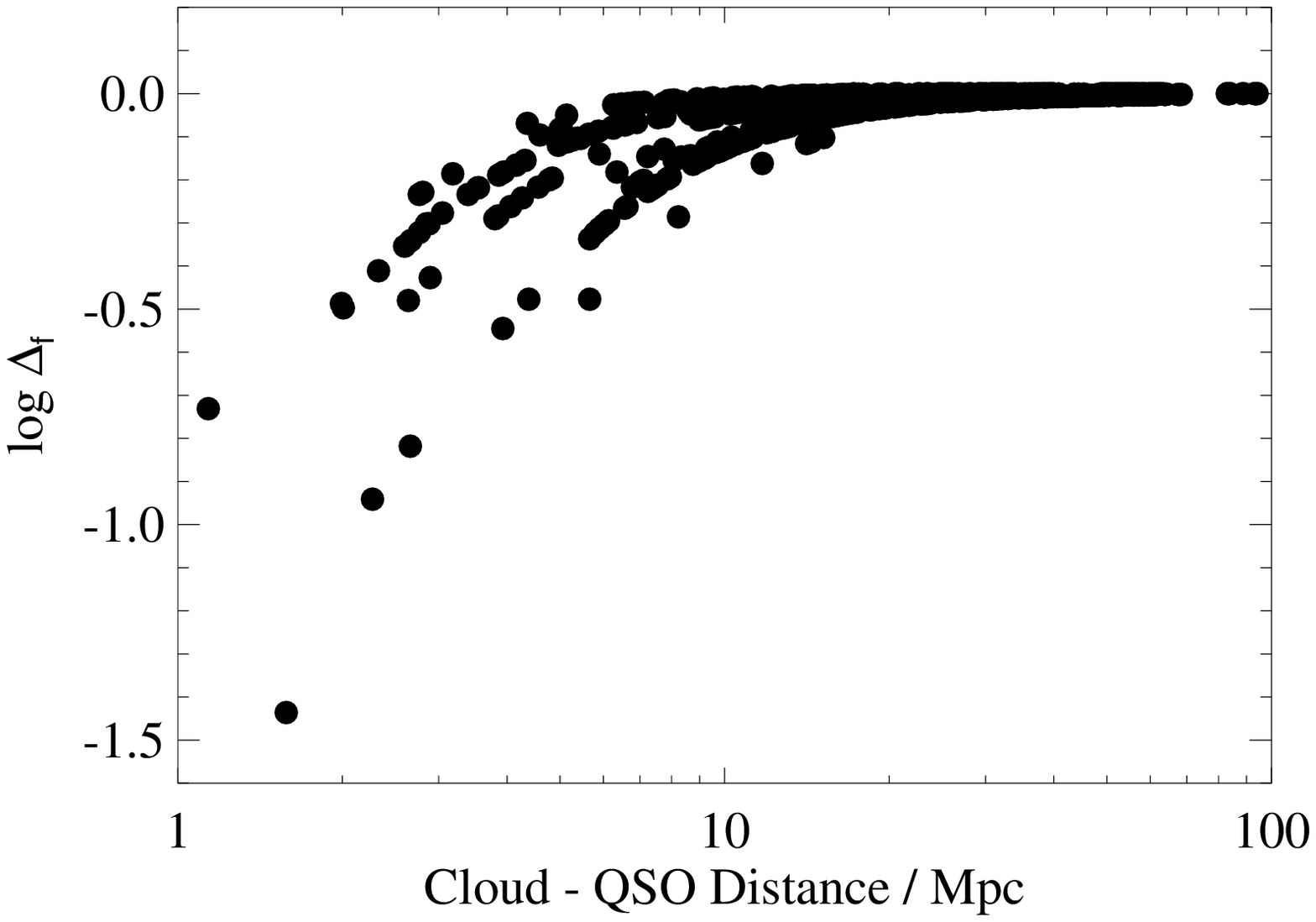}
}
\epsfverbosetrue
\caption{On the left, absorber redshifts are plotted against $\Delta_F$.  On the right $\Delta_F$ is plotted against the distance between cloud and quasar.  Note that the correction for the quasar's flux, and hence the upper limit to the estimate of the background, is significant for only a small fraction of the clouds.}
\label{fig:prox}
\end{figure*}

While there is little evidence here for evolution of the background,
the upper limits to the background flux diverge more strongly than the
lower limits at the lowest and highest redshifts.  Also, the posterior
probability of the background is extended more towards higher values.

The background was measured by comparing its effect with that of the
quasar.  If the background were larger the quasar would have less
effect and the clouds with $\Delta_F < 1$ would not need as large a
correction to the observed column density for them to agree with the
population as a whole.  If the background was less strong then the
quasars would have a stronger influence and more clouds would be
affected.

The upper limit to the flux depends on clouds influenced by the
quasar.  Figure \ref{fig:prox} shows how $\Delta_F$ changes with
redshift and proximity to the background quasar.  From this figure it
is clear that the upper limit is dominated by only a few clouds.
However, the lower limit also depends on clouds near to, but not
influenced by, the quasar.  This involves many more clouds.  The lower
limit is therefore stronger, more uniform, and less sensitive to the
amount of data, than the upper limit.

Other procedures for calculating the errors in the flux have assumed
that the error is symmetrical (the only apparent exception is
Fern\'{a}ndez--Soto et~al. (1995) who unfortunately had insufficient data to normalize
the distribution).  While this is acceptable for $\beta$ and $\gamma$,
whose posterior probability distributions
(figure~\ref{fig:beta_gamma_d}) can be well--approximated by Gaussian
curves, it is clearly wrong for the background
(eg. figure~\ref{fig:flux_both}), especially where there are less data
(at the lowest and highest redshifts).

An estimate based on the assumption that the error is normally
distributed will be biased in two ways.  First, since the extended
upper bound to the background has been ignored, it will underestimate
the `best' value.  Second, since the error bounds are calculated from
the curvature of the posterior distribution at its peak (ie. from the
Hessian matrix) they will not take account of the extended `tails' and
so will underestimate the true range of values.  In addition, most
earlier work has calculated errors assuming that the other population
parameters are fixed at their best--fit values.  This will also
under--estimate the true error limits.  All these biases become more
significant as the amount of data decreases.

\begin{figure*}
\epsfxsize=15.cm
\epsfbox{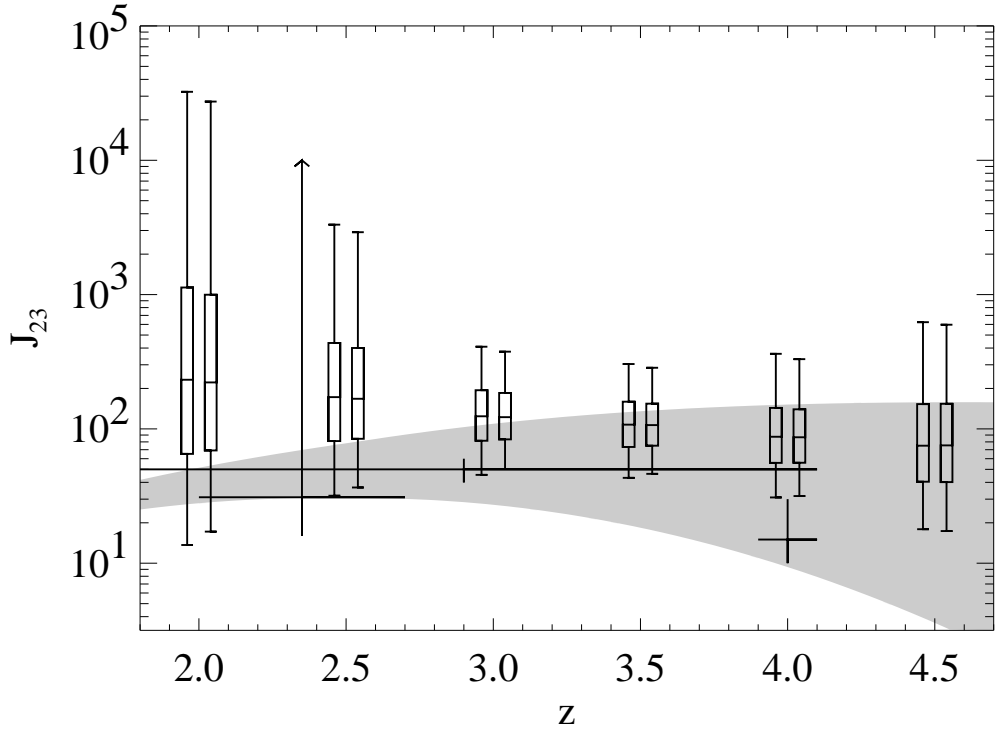}
\epsfverbosetrue

\caption{The expected probability distribution of the log background
flux for models {\bf D} (left) and {\bf E} (right, including a
correction for the known damped absorption systems).  The box plots
show median, quartiles, and 95\% limits.  The shaded area covers the
range of backgrounds described in Fall \& Pei (1995).  The lower boundary
is the expected background if all quasars are visible, the higher
fluxes are possible if an increasing fraction of the quasar population
is obscured at higher redshifts.  The crosses and arrows mark the
extent of previous measurements from high resolution spectra --- see
the text for more details.}

\label{fig:evoln}
\end{figure*}

The first of these biases also makes the interpretation of the
box--plots (eg. figures \ref{fig:flux_both} and \ref{fig:evoln}) more
difficult.  For example, the curves in the left--hand plot in
figure~\ref{fig:flux_both} and the data in table~\ref{tab:modeld} show
that the value of the flux with highest probability at $z=2$ is
$140$~\jflux\ (for model {\bf D}).  In contrast the box--plot on the
right shows that the median probability is almost twice as large
($230$~\jflux).  Neither plot is `wrong': this is the consequence of
asymmetric error distributions.

%
%

\subsection{Comparison with Previous Estimates}

%
%

\subsubsection{Earlier High--Resolution Work}
\label{sec:prevhi}

Fern\'{a}ndez--Soto et~al. (1995) fitted high signal--to--noise data towards three
quasars.  For $2 < z < 2.7$\ they estimate an ionizing background
intensity of 32~\jflux, with an absolute lower limit (95\% confidence)
of 16~\jflux\ (figure~\ref{fig:evoln}, the leftmost cross). They were
unable to put any upper limit on their results.

Cristiani et~al. (1995) determined a value of 50~\jflux\ using a sample of five
quasars with a lower column density cut--off of log($N$) = 13.3.  This
sample was recently extended (Giallongo 1995).  They find that the
ionizing background is roughly constant over the range $1.7 < z <
4.0$\ with a value of 50~\jflux\ which they considered a possible
lower limit (figure~\ref{fig:evoln}, the middle lower limit).

While this paper was being refereed Giallongo et~al. (1996) became available,
extending the work above.  Using a maximum likelihood analysis with an
unspecified procedure for calculating errors they give an estimate for
the background of $50\pm10$~\jflux.  They found no evidence for
evolution with redshift when using a single power law exponent.

Williger et~al. (1994) used a single object, Q1033--033, which is included in
this sample, to give an estimate of $10-30$~\jflux\
(figure~\ref{fig:evoln}, the rightmost cross).  The error limits are
smaller than those found here, even though they only use a subset of
this data, which suggests that they have been significantly
underestimated.

If the errors in Williger et~al. (1994) are indeed underestimates then these
measurements are consistent with the results here.  However, the
best--fit values are all lower than those found here.  This may be, at
least partly, because of the biases discussed in
section~\ref{sec:lims}.  

Williger et~al. (1994) used a more direct method than usual to estimate the
background.  This gives a useful constraint on the effect of
line blending in the procedures used, and is explored in more detail
below.

%
%

\subsubsection{Q1033--033 and Line Blending\label{sec:gerry}}

The measured value of the background, 80\elim{80}{40}~\jflux\ (model
{\bf D} at $z=4$), is larger than an earlier estimate using a subset
of this data (Williger et~al. 1994, Q1033--033, $10-30$~\jflux).

As has already been argued, it is difficult to understand how a
procedure using much less data could have smaller error limits than
the results here, so it is likely that the error was an underestimate
and that the two results are consistent.  However, it is interesting
to see if there is also a systematic bias in the analyses used.

The correction for galactic absorption is not very large for this
object (about 20\%).  More importantly, the procedures used differ
significantly in how they are affected by blended lines.  These are a
problem at the highest redshifts, where the increased \lya\ cloud
population density means that it is not always possible to resolve
individual clouds.  Williger et~al. (1994) added additional lines ($\col =
13.7$, $b$ = 30~\kms) to their $z$ = 4.26 spectra of Q1033--033 and
found that between 40\% and 75\% would be missed in the line list.

As the lower column density limit is raised Williger et~al. (1994) find that
the observed value of $\gamma$ also increases.  The resulting stronger
redshift evolution would make the deficit of clouds near the quasar
more significant and so give a lower estimate of the background.
Although not significant at the 95\% level, there is an indication
that $\gamma$ also increases with higher column density in this
analysis (section \ref{sec:dispop}, figure \ref{fig:var_d}).  While it
is possible that $\gamma$ varies with column density the same
dependence would be expected if line blending is reducing the number
of smaller clouds. To understand how line blending can affect the
estimates, we will now examine the two analyses in more detail.

Line blending makes the detection of lines less likely.  Near the
quasar lines are easier to detect because the forest is more sparse.
In the analysis used in this paper the appearance of these `extra'
lines reduces the apparent effect of the quasar.  Alternatively, one
can say that away from the quasar line blending lowers $\gamma$.
Both arguments describe the same process and imply that the estimated
background flux is too large.

In contrast, Williger et~al. (1994) take a line--list from a crowded region,
which has too few weak lines and correspondingly more saturated lines,
and reduce the column densities until they agree with a region closer
to the quasar.  Since a few saturated lines are less sensitive to the
quasar's flux than a larger number of weaker lines, the effect of this
flux is over--estimated (and poorly determined), making the background
seem less significant and giving a final value for the background flux
which is too small.  This method is therefore biased in the opposite
sense to ours and so the true value of the background probably lies
between their estimate and ours.

The comparison with Williger et~al. (1994) gives one estimate of the bias from
line blending.  Another can be made by raising the completeness limits
of the data (section~\ref{sec:malm}).  This should decrease the number
of weak, blended lines, but also excludes approximately half the data.
In figure \ref{fig:flux_both} the flux estimates from the full data
set are shown together with those from one in which the limits have
been raised by $\Delta\log(N)=0.5$.  There is little change in the
lowest reasonable flux, an increase in the upper limits, and in
increase in the `best--fit' values.  The flux for $z<3$ is almost
unconstrained by the restricted sample (section~\ref{sec:lims}
explains the asymmetry).

An increase of 0.5 in $\log(N)$ is a substantial change in the
completeness limits.  That the lower limits remain constant (to within
$\sim 0.1$ dex) suggests that line blending is not causing the flux to
be significantly over--estimated.  The increase in the upper limits is
expected when the number of clouds in the sample decreases
(section~\ref{sec:lims}).

In summary, the total difference between our measurement and that in
Williger et~al. (1994) is 0.7 dex which can be taken as an upper limit on the
effect of line blending.  However, a more typical value, from the
constancy of the lower limits when completeness limits are raised, is
probably $\sim0.1$ dex.

%
%

\subsubsection{Results from Lower Resolution Spectra}

Bechtold (1994) analysed lower resolution spectra towards 34 quasars
using equivalent widths rather than individual column density
measurements.  She derived a background flux of 300~\jflux\
($1.6<z<4.1$), decreasing to 100~\jflux\ when a uniform correction was
applied to correct for non--systemic quasar redshifts.  With
low--resolution data a value of $\beta$ is used to change from a
distribution of equivalent widths to column densities.  If $\beta$ is
decreased from 1.7 to a value closer to that found for narrower lines
(see section~\ref{sec:dispop}) then the inferred background estimate
could decrease further.

The evolution was not well--constrained ($-7<\alpha<4$).  No
distinction was made between the lower and upper constraints on the
flux estimate, and it is likely that the wide range of values reflects
the lack of strong upper constraints which we see in our analysis.

It is not clear to what extent this analysis is affected by line
blending.  Certainly the comments above --- that relatively more
clouds will be detected near the quasar --- also apply.

%
%

\subsubsection{Lower Redshift Measurements}

The background intensity presented in this paper is much larger than
the 8~\jflux\ upper limit at $z=0$ found by
Vogel et~al. (1995). Kulkarni \& Fall (1993) obtain an even lower value of
0.6\elim{1.2}{0.4}~\jflux\ at $z=0.5$ by analysing the proximity
effect in HST observations.  However, even an unevolving flux will
decrease by a factor of $\sim 50$ between $z=2$ and $0$, so such a
decline is not inconsistent with the results given here.

%
%

\subsection{What is the Source of the Background?}
\label{sec:source}

\subsubsection{Quasars}

Quasars are likely to provide a significant, if not dominant,
contribution to the extragalactic background.  An estimate of the
ionizing background can be calculated from models of the quasar
population.  Figure \ref{fig:evoln} shows the constraints from models
{\bf D} and {\bf E} and compares them with the expected evolution of
the background calculated by Fall \& Pei (1995).  The background can take
a range of values (the shaded region), with the lower boundary
indicating the expected trend for a dust--free universe and larger
values taking into account those quasars that may be hidden from our
view, but which still contribute to the intergalactic ionizing flux.
The hypothesis that the flux is only from visible quasars (the
unobscured model in Fall \& Pei 1995) is formally rejected at over the
95\% significance level since the predicted evolution is outside the
95\% bar in the box plots at higher redshift.

Although our background estimate excludes a simple quasar--dominated
model based on the observed number of such objects, the analysis here
may give a background flux which is biased (too large) from a
combination of line blending (section~\ref{sec:gerry}) and clustering
around the background quasars.  From the comparison with
Williger et~al. (1994), above, there is an upper limit on the correction for
line blending, at the higher redshifts, of 0.7 dex.  However, an
analysis of the data when column density completeness limits were
increased by $\Delta\log(N)=0.5$ suggests that a change in the lower
limits here of $\sim 0.1$ dex is more likely.  A further change of up
to between 0.5 and 1 dex is possible if quasars lie in regions of
increased clustering (section~\ref{sec:syserr}).  These two effects
imply that at the highest redshifts the flux measured here could
reasonably overestimate the real value by $\sim 0.5$ dex.  This could
make the measurements marginally consistent with the expected flux
from the observed population of quasars.

There is also some uncertainty in the expected background from quasars
since observations could be incomplete even at the better understood
lower redshifts (eg.~Goldschmidt et~al. 1992) and while absorption in damped
systems is understood in theory (Fall \& Pei 1993) its effect is
uncertain (particularly because the distribution of high column
density systems is poorly constrained).  

The highest flux model (largest population of obscured quasars) from
Fall \& Pei (1995) is consistent with the measurements here (assuming that
the objects used in this paper are not significantly obscured).

\subsubsection{Stars}

The background appears to be stronger than the integrated flux from
the known quasar population.  Can star formation at high redshifts
explain the discrepancy?  

Recent results from observations of low redshift starbursts
(Leitherer et~al. 1995) suggest that very few ionizing photons ($\leq 3\,$\%)
escape from these systems.  If high redshift starbursts are similar in
their properties, then the presence of cool gas in these objects would
similarly limit their contribution to the ionizing background.
However, Madau \& Shull (1996) estimate that if star formation occurs in
\lya\ clouds, and a significant fraction of the ionizing photons
($\sim 25\,\%$) escape, then these photons may contribute a
substantial fraction of the ionizing background photons in their
immediate vicinity.  As an example, at $z \sim 3$\ they estimate that
$J_\nu \leq 50~\hbox{\jflux}$\ if star formation sets in at
$z\sim3.2$.  This flux would dominate the lowest (no correction for
obscuration) quasar background shown in figure \ref{fig:evoln} and
could be consistent with the intensity we estimate for the background
at this redshift, given the possible systematic biases discussed above
and in section~\ref{sec:syserr}.

%
%

\section{Conclusions}
\label{sec:conc}

A model has been fitted to the population of \lya\ clouds.  The model
includes the relative effect of the ionising flux from the background
and nearby quasars (section~\ref{sec:model}).

The derived model parameters for the population of absorbers are
generally consistent with earlier estimates.  There is some evidence
that $\beta$, the column density power law population exponent,
increases with column density, but could also be due to line blending
(section~\ref{sec:dispop}).

The ionising background is estimated to be 100\elim{50}{30}~\jflux\
(model {\bf B}, section~\ref{sec:resbackg}) over the range of
redshifts ($2<z<4.5$) covered by the data.  No strong evidence for
evolution in the ionizing background is seen over this redshift range.
In particular, there is no significant evidence for a decline for
$z>3$ (section~\ref{sec:noevoln}).  Previous results may have been
biased (too low, with optimistic error limits ---
section~\ref{sec:lims}).

Constraints on the evolution of the background are shown in figure
\ref{fig:evoln}.  The estimates are not consistent with the background
flux expected from the observed population of quasars
(section~\ref{sec:source}).  However, two effects are likely to be
important.  First, both line blending and increased clustering of
clouds near quasars lead to the measured background being
overestimated.  Second, a significant fraction of the quasar
population at high redshifts may be obscured.  Since their
contribution to the background would then be underestimated this would
imply that current models of the ionizing background are too low.
Both of these would bring the expected and measured fluxes into closer
agreement.  It is also possible that gravitational lensing makes the
measurement here an overestimate of the true background.

The dominant source of errors in our work is the limited number of
lines near the background quasar (eg. figures \ref{fig:nu1_d}\ and
\ref{fig:prox}).  Systematic errors are smaller and become important
only if it is necessary to make standard (unobscured quasar) models
for the background consistent with the lower limits presented here.
Further data will therefore make the estimate here more accurate,
although observational data are limited by confusion of the most
numerous lower column density systems ($\col< 13.0$) so it will
remain difficult to remove the bias from line blending.  An
improvement in the errors for the highest redshift data points, or a
determination of the shape of the ionizing spectrum (e.g. from
He~II/H~I estimates in \lya\ clouds) would help in discriminating
between current competing models for the ionizing background.
Finally, a determination of the background strength in the redshift
range $0.5 < z < 2.0$ is still needed.

%
%

\section{Acknowledgements}

We would like to thank Yichuan Pei for stimulating discussions and for
making data available to us.  Tom Leonard (Dept. of Statistics,
Edinburgh) gave useful comments and guidance on the statistics used in
this paper.  We would also like to thank an anonymous referee for
helpful and constructive comments.

\end{document}